\documentclass[english]{article}
\usepackage{jcappub}
\usepackage{lmodern}

\usepackage{graphicx}
\usepackage{esint}
\usepackage{babel}

\def\lsim{\mathrel{\rlap{\lower4pt\hbox{\hskip1pt$\sim$}}
    \raise1pt\hbox{$<$}}}                
\def\gsim{\mathrel{\rlap{\lower4pt\hbox{\hskip1pt$\sim$}}
    \raise1pt\hbox{$>$}}}                

\makeatletter

\begin{document}
\title{Estimating SI violation in CMB due to non-circular beam and complex scan in minutes}

\author[a]{Nidhi Pant}
\author[a]{Santanu Das}
\author[b]{Aditya Rotti}
\author[a]{Sanjit Mitra}
\author[a]{Tarun Souradeep}

\emailAdd{nidhip@iucaa.in}
\emailAdd{santanud@iucaa.ernet.in}
\emailAdd{arotti@fsu.edu}
\emailAdd{sanjit@iucaa.in}
\emailAdd{tarun@iucaa.in}

\affiliation[a]{IUCAA, Post Bag 4, Ganeshkhind, Pune-411007, India}
\affiliation[b]{Department of Physics, Florida State University, Tallahassee, FL 32304, USA}

\abstract{
Mild, unavoidable deviations from circular-symmetry of instrumental
beams along with scan strategy can give rise to measurable Statistical Isotropy (SI) violation 
in Cosmic Microwave Background (CMB) experiments. If not accounted properly, 
this spurious signal can complicate the extraction of other SI violation signals (if any) in the data. 
However, estimation of this effect through exact numerical simulation is computationally intensive and time consuming. 
A generalized analytical formalism not only provides a quick way of estimating this signal, 
but also gives a detailed understanding connecting the leading beam anisotropy components to a 
measurable BipoSH characterisation of SI violation. 
In this paper, we provide an approximate generic analytical method for estimating the SI violation
generated due to a non-circular (NC) beam and arbitrary scan strategy, in terms of the Bipolar
Spherical Harmonic (BipoSH) spectra. 
Our analytical method can predict almost all the features introduced by a NC beam in a complex scan and thus reduces 
the need for extensive numerical simulation worth tens of thousands of CPU hours into minutes long calculations. 
 As an illustrative example, we use WMAP beams and scanning strategy 
to demonstrate the easability, usability and efficiency of our method. We test all our analytical results against that from exact 
numerical simulations. 
}
\maketitle

\section{Introduction}

Observed CMB anisotropy on the sky is a
convolution of the underlying cosmological CMB signal with the
instrumental beam response function.  The instrumental beam (response function) in most
CMB experiments are designed to be nearly circularly (azimuthal) symmetric. However, mild
deviations from circularity do inevitably arise due to unavoidable limitations in experimental 
design, function and fabrication; e.g., the primary lobe of the beam
exhibits non-circularity due to the off-axis position of detectors on
the focal plane; diffraction around the edges of instrument leads to
side lobes of the beam; or due to finite response time of detectors, 
the scan may not correspond to the direction of 
its beam axis leading to the effective beam response at any pointing direction
being sensitive to the scan strategy, etc.
Regardless of the specific origin of non-circularity, beam imperfections, coupled with the scan
strategy lead to very complex modification of the signal demanding
high computational resources to assess the final effect on the
estimation of angular power spectrum, cosmological
parameters and Statistical Isotropy~(SI) violation etc.

Cosmological CMB temperature fluctuations are generally assumed to be a realization of statistically isotropic,
Gaussian, correlated random field on the sphere. Consequently, the
angular power spectrum has been the primary observational target of
most CMB experiments.
The effect of NC beam on angular power spectrum of CMB
has been studied in literature and the non-trivial impact on high
precision cosmological inferences has been appreciated but not
satisfactorily resolved, particularly, within the available computational
resources. A significant body of literature attempting to deal with NC
beam effect on the angular power spectrum exists, e.g.,
~\cite{TS-BR,PF-OD-FB,SM-AS-TS,TS-SM-AS-SR-RS,FebeCoP2011,fastconv2001,WMAP_GH2006,Planckmap_Ash2011,Das2015T,SD-TS2012a,Das:2012ft}.
However, current and upcoming CMB experiments also hold the promise to observationally constrain the underlying,
often implicit, SI assumption (closely linked to the so called, `cosmological principle'), which implies rotational invariance of the $n$-point correlation function.
SI assumption has been under intense scrutiny with hints of various `anomalies'
persisting in successive years of WMAP
data and recently Planck data~\cite{MT-OC-AH,PB-KG-AB,CC-DH-DS,KL-JM,HE-FH-AB,Planck-SI,Planck15,Das2014,Das2015,Mukherjee2014,Mukherjee2013a}.

Violation of SI can arise both from theoretical possibilities or from observational artefacts~\cite{ML-JL,NC-DS-GS,BPS2000,JL,LA-SM-MB,AG-CC-MP,TS2006,MA-TS,AP-MK,AR-MA-TS,Mukherjee2013}.
Cosmic topology, anisotropic cosmologies, Doppler boost etc.  give some of the theoretical source 
for SI violation. On the other hand 
observational artefacts include beam non-circularity, anisotropic
noise, foreground residuals, masking etc.
Whatever be the source of the SI violation, breakdown of SI can be parametrized by expanding the two point correlation function in 
the Bipolar Spherical Harmonic (BipoSH) basis~\cite{AH-TS-03}. This parametrization captures the SI violation in a 
mathematically structured representation. NC-beam along with complex scan strategy induces SI violation in an otherwise 
SI sky and thus pose as a serious systematic contaminant in SI measurements. 
We use BipoSH basis to characterize the effect. 

In this paper, we expand the NC-beam functions 
in BipoSH basis and the coefficients of expansion are referred as {\em beam-BipoSH} coefficients ($B^{LM}_{l_1 l_2}$). An ideal 
circularly symmetric beam only have $L=0$ non-vanishing beam-BipoSH coefficients. Breakdown of circular symmetry further induces $L \neq 0$ modes in the beam-BipoSH coefficients.
Most NC beams have mild deviation from circular symmetry, which reflects as a dominant $m=0$ mode in the beam spherical harmonic coefficients, $b_{lm}$.
In most of the realistic beams, $b_{lm}/b_{l0}$ decreases rapidly with increasing m for each $l$. Since, any realistic beam has dominant even-fold symmetry, the odd $m$ modes are also negligible.   We provide simple explicit analytic expressions for beam-BipoSH coefficients for any experimental beam and scan. 
We show that NC-beam introduces the SI violation signals in the measured sky-map such that every non-zero beam-BipoSH coefficient ($B^{LM}_{l_1 l_2}$) generates a corresponding non-zero CMB-BipoSH coefficient ($A^{LM}_{l_1 l_2}$), and  analytically relate the beam-BipoSH coefficients with the CMB-BipoSH coefficients for a generalized scan.

Numerical estimation of the BipoSH spectra generated due to experimental beam and scan 
requires generating multiple realizations of beam convolved CMB maps with the given scan strategy.
First step to generate each such realization requires generation of the time order data (TOD) by 
convolving the random SI sky (generated by HEALPix~\cite{hpix}) with the experimental beam in each time step along the scan path.
Thereafter, map-making is used to obtain the realizations from the TOD. 
This entire process is highly time consuming and computationally expensive. 
However, our approximate semi-analytic formalism to estimate the effect of mildly NC beam 
with the experimental scan on the observed CMB-BipoSH coefficients is fast and gives an insight to the 
corresponding characteristic form of the SI violation.
We verify all our analytical results 
with the results from exact extensive numerical simulations. 

The paper is organized as follows. Sec.~\ref{bipfor} provides a brief
primer to the BipoSH formalism to characterize SI violations for keeping
the paper self contained. In Sec.~\ref{ncbeambposh}, we present a novel
expansion of the beam response function in the BipoSH basis. 
According to our formulation, the beam-BipoSH coefficients 
depend on the scan pattern. Therefore, in Sec.~\ref{Sec:PTscan} we provide a method 
for evaluating the beam-BipoSH coefficients in a simplified scan pattern, 
referred as parallel transport (PT) scan. We provide the detailed 
expressions for the beam-BipoSH coefficients in PT scan coordinates. 
In Sec.~\ref{bipforbeam}, we derive expressions for the CMB-BipoSH
coefficients, arising due to convolution of SI sky-map 
with a NC-beam, in terms of the beam-BipoSH coefficients. 
In Sec.~\ref{app:EG} and Sec.~\ref{BipoSHWMAPrawbeam}, we validate our analytical results against numerical simulations for elliptical Gaussian beam and the WMAP raw beam respectively in a PT scan. 
Next, in Sec.~\ref{Sec:WMAPapp} we evaluate the expressions for the CMB-BipoSH coefficients for a generalized scan with NC beam in terms of 
the CMB-BipoSH coefficients with same beam but PT scan. All these analytical results are verified with numerical simulations. 
Sec.~\ref{conclusions} presents the discussions and conclusions of this paper. Detailed steps
of all analytical calculations are provided for completeness in
Appendix~\ref{app:cmb-biposh} and Appendix~\ref{app:beam-biposh}. 

\section{Primer: Bipolar Spherical Harmonic (${\rm  BipoSH}$) representation}\label{bipfor}
Statistical Isotropy (SI) implies rotational invariance of $N$-point correlation function and 
enforces the two point correlation function $C(\hat n_{1}, \hat n_{2})$ to be only a function of the angular separation $(\hat n_{1}\cdot \hat n_{2})$. Consequently, $C(\hat n_{1}\cdot \hat n_{2})$
It can be expanded in terms of Legendre polynomials where coefficients of expansion are well known CMB angular power spectrum, $C_l$. In harmonic space, this condition translates to diagonal covariance matrix,

\begin{eqnarray}\label{harmonic-SI}
\langle a_{lm} a^{*}_{l'm'}\rangle=C_{l}\delta_{ll'}\delta_{mm'},
\label{SI_SH_cov}
\end{eqnarray}

\noindent where $a_{lm}$'s are spherical harmonic coefficients of expansion of CMB temperature field, $\Delta T(\hat n)$ and the 
angular bracket denotes the ensemble average.
Eq.(\ref{harmonic-SI}) implies that (the $m$-independent), $C_l$ encodes all the
information in a SI field on the full sky (complete sphere,
$\textbf{S}^{2}$). 

However, in presence of SI violation the covariance matrix, $\langle a_{lm} a^{*}_{l'm'}\rangle$ will, in general, have additional terms in the diagonal beyond 
$C_{l}\delta_{ll'}\delta_{mm'}$ and also the off-diagonal components. 
The two point correlation function, then depends on both the
directions $\hat n_1$ and $\hat {n_2}$ and not just on the angle
between them and most generally can be expanded in Bipolar
Spherical Harmonic(BipoSH)
basis~\cite{AH-TS-03,AH-TS-04,AH-TS-NC,AH-TS-05,SB-AH-TS,AH-TS-06} as

 \begin{equation}\label{eq:BPOSH}
C(\hat{n}_{1},\hat{n}_{2}) =
\sum_{l_{1},l_{2},L,M}A_{l_{1}l_{2}}^{L
M}\{Y_{l_{1}}(\hat{n}_{1})\otimes Y_{l_{2}}(\hat{n}_{2})\}_{L M},
\end{equation}

\noindent where $A_{l_{1}l_{2}}^{L M}$ are called the BipoSH coefficients. The bipolar
spherical harmonic (BipoSH) functions,

\begin{eqnarray}\label{eq:BipoSH}
\{Y_{l_{1}}(\hat{n}_{1})\otimes Y_{l_{2}}(\hat{n}_{2})\}_{L M}= \sum_{m_{1}m_{2}} C_{l_{1}m_{1}l_{2}m_{2}}^{LM}Y_{l_{1} m_{1}}(\hat{n}_{1})\;
Y_{l_{2}m_{2}}(\hat{n}_{2})\,,
\end{eqnarray}

\noindent are irreducible tensor product of two spherical harmonics spaces that
form an orthonormal basis on $\textbf{S}^{2} \times \textbf{S}^{2}$. 
$C_{l_{1}m_{1}l_{2}m_{2}}^{LM}$ are the Clebsch-Gordon
coefficients. The multipole indices  of these coefficients satisfy the
triangularity conditions $|l_1-l_2|\le L\le l_1+l_2$ and $m_1+m_2=M$.

We can show that the BipoSH coefficients are given by~\cite{AH-TS-03},
\begin{eqnarray}\label{eq:gen-BipoSH}
A^{LM}_{l_1 l_2}=\sum_{m_1 m_2}\langle a_{l_1 m_1}a^{*}_{l_2
m_2}\rangle(-1)^{m_2}C^{LM}_{l_1 m_1 l_2 -m_2}.
\end{eqnarray}

The BipoSH coefficients $A^{00}_{ll}$ in  Eq.(\ref{eq:BPOSH}) corresponds to the SI part 
and can be expressed in terms of the CMB angular power spectrum as $A^{00}_{ll}=(-1)^{l} C_{l}\prod_{l}$, where
$\prod_{ab...c}=\sqrt{(2a+1)(2b+1)...(2c+1)}$ \cite{AH-TS-03}.  

{\em Non-zero BipoSH coefficients with
$L>0$ capture SI violation
}~\cite{NJ-SJ-TS-AH2010}. 
The BipoSH coefficients can be categorized into two distinct classes, defined as even ( $l_1+l_2+L$ is even) and odd ($l_1+l_2+L$ is odd) parity BipoSH.
This distinction provides valuable clues to the origin of SI violations e.g., weak lensing due
to scalar (even parity) and tensor (odd parity)
perturbations~\cite{LB-MK-TS2012}, anisotropic primordial power
spectrum (even)~\cite{AP-MK}, temperature modulation
(even)~\cite{DH-AL}, primordial homogeneous magnetic fields
(even)~\cite{AH-PhD,MA-TS}. Importantly, in the context of NC-beam effect, the absence of
significant odd parity BipoSH would imply a reflection symmetric NC-beam.

\section{${\rm Beam-BipoSH}$: Non-circular beams in ${\rm BipoSH}$ representation}
\label{ncbeambposh}

Beam function about
the pointing direction $\hat n$ can be decomposed in Spherical Harmonic (SH) basis as,
\begin{eqnarray}
B(\hat n,\hat n_{1})=\sum_{lm}b_{lm}(\hat n)Y_{lm}(\hat n_{1}).
\label{beamSHexp}
\end{eqnarray}
The SH transform of beam at arbitrary pointing direction, $\hat
n\equiv(\theta,\phi)$ is given by rotating the {\em beam-SH}, $b_{l m'}(\hat z)$ --
the SH transform of the beam pointing along fixed direction $\hat z$,
\begin{eqnarray}\label{eq:b2}
b_{lm}(\hat n)=\sum_{m'}b_{l m'}(\hat z) D^{l}_{m
m'}(\phi,\theta,\rho(\hat n)),
\label{beamSHz}
\end{eqnarray}
where Wigner D-functions $D^{l}_{mm'}(\alpha,\beta,\gamma)$, are the
matrix elements of the rotation operator ($0\leq\alpha<2\pi,\
0\leq\beta<\pi,\ 0\leq\gamma<2\pi$) and $\alpha, \beta, \gamma$ are
the Euler angles that rotate the $\hat z$-axis to the pointing
direction $\hat n=(\theta,\phi)$ and the angle $\rho(\hat n)$ specifies
the orientation of the NC-beam with respect to the local Cartesian coordinates
$(\hat x\equiv\hat\phi, \hat y\equiv\hat\theta)$~\cite{TS-BR}. 
Such a rotation can be realized
by fixing a coordinate system and performing anti-clockwise rotations,
first rotating about the $\hat z$-axis by an angle $\alpha=\phi$, then
rotating about new $\hat y$-axis by an angle $\beta=\theta$, and
finally about the new $\hat z$-axis by $\gamma=\rho(\hat n)$.

Since, a general NC-Beam function depends
on two vector directions, it can be expanded in the BipoSH basis (see
Sec.~\ref{bipfor}),
\begin{eqnarray}
B(\hat n,\hat n_{1}) \ = \label{eq:beambipolar}
 \sum_{l_1 l_2 L M} B^{LM}_{l_1 l_2}
\sum_{m_1 m_2} C^{LM}_{l_1 m_1 l_2 m_2} Y_{l_1 m_1}(\hat n) Y_{l_2
m_2}(\hat n_{1}), 
\end{eqnarray}
where the coefficients of expansion $B^{LM}_{l_1 l_2}$ are referred to
as {\em beam-BipoSH} coefficients . 

The beam-BipoSH coefficient $B^{LM}_{l_1 l_2}$, can be readily related to beam-SH coefficients $b_{l m}(\hat n)$  as
\begin{equation}\label{eq:b1}
B^{LM}_{l_1 l_2}=\sum_{m_1 m_2}C^{LM}_{l_1 m_1 l_2 m_2} \int
d\Omega_{\hat n} b_{l_2 m_2}({\hat n})Y^{*}_{l_1 m_1}(\hat n)\,.
\end{equation}
A circularly symmetric beam function around the pointing direction can
be expanded in Legendre polynomials, $B(\hat n,\hat n_{1})\equiv B(\hat
n\cdot\hat n_{1})={(4\pi)}^{-1}\sum_l (2l+1)B_{l}P_{l}(\hat
n\cdot\hat n_{1})$.  Inverse transforming
Eq.(\ref{eq:beambipolar}) and using orthogonality of
BipoSH~\cite{varshalovich}, we obtain beam-BipoSH coefficients for
circularly symmetric beam function,
\begin{eqnarray}\label{beamcircular}
B^{LM}_{l_1 l_2}=(-1)^{l_1}B_{l_1}\prod_{l_1}\delta_{l_1 l_2}\delta_{L0}\delta_{M0},
\end{eqnarray}
where $B_{l}$ is the commonly used Legendre transform of the beam
function in the circularized beam approximation.

Beam-BipoSH depend not only on NC-beam
harmonics but also on the scan-strategy that defines $\rho(\hat n)$, at arbitrary pointing direction, $\hat
n\equiv(\theta,\phi)$.
For any {\em arbitrary scanning strategy}, using Eq.(\ref{eq:b2}) and Eq.(\ref{eq:b1}), 
it turns out that the beam-BipoSH can be expressed in terms of the beam-SH and scanning parameter $\rho(\hat n)$ as
\begin{eqnarray}\label{eq:beam-BipoSH}
B^{LM}_{l_1 l_2} \ =\ \sum_{m'}b_{l_2
m'}(\hat z)\Big(\sum_{m_1 m_2}C^{LM}_{l_1 m_1 l_2 m_2} \ \times \label{eq:genbeambiposh1}
\quad \int^{\pi}_{0}\int^{2\pi}_0  D^{l_2}_{m_2
m'}(\phi,\theta,\rho(\hat n))Y^{*}_{l_1
m_1}(\hat n)\sin\theta~ d\theta~ d\phi\Big). 
\end{eqnarray}
To separate the azimuthal ($\phi$) and polar ($\theta$) dependencies,
it is convenient to express Wigner-\textit{D} functions in terms of
Wigner-\textit{d} through following relation,
\begin{eqnarray}\label{eq:wigner-scan}
D^{l}_{m m'}(\phi,\theta,\rho(\hat n))={\rm e}^{-i m \phi}\,d^{l}_{m
m'}(\theta)\, {\rm e}^{-i m' \rho(\hat n)}.
\end{eqnarray}

\noindent Eq.(\ref{eq:genbeambiposh1}) is the most general expression of
beam-BipoSH coefficients for single hit for any given NC-beam specified through
$b_{lm}(\hat z)$ and scan pattern, defined by $\rho(\theta,\phi)$, in
any spherical polar coordinate system (e.g. ecliptic, galactic, etc.). 

Analytic progress to evaluate beam-BipoSH coefficients is less tedious when the beam has mild deviations from
circularity and allows to retain only the leading order terms up to $|m'|=2$ of the beam-SH.
Further, in most realistic beam, the beam function has a dominant even fold azimuthal symmetry such that only even values of $m'$ is allowed.
Hence, throughout the rest of the paper we have truncated the summation over $m'$ in Eq.(\ref{eq:beam-BipoSH}) with $m'=0, \pm 2$.
It is worthy to note that $m'=0$ is the circular part of the beam. Non-circularity of the beam is characterized by $m'=2$.
In BipoSH space, the consequence of discrete even-fold azimuthal and
reflection symmetric NC-beam translates to restricting non-zero
beam-BipoSH to $M=\rm{even}$ and $l_{1}+l_{2}=\textrm{even}$ respectively.

\subsection{Beam-BipoSH in `Parallel-transport' scan approximation}\label{Sec:PTscan}
The general beam-BipoSH in Eq.(\ref{eq:genbeambiposh1}) can be tackled
analytically when the scan pattern is such that $\rho(\hat n)$ is a
constant. We refer such a scan pattern as {\em `parallel-transport' (PT) scan}
following~\cite{TS-BR}.  It implies that the
orientation of the beam relative to the local longitude is constant at
any point on the sky.  
Note that a constant $\rho$ can be absorbed as phase factor in the redefinition of
the complex quantity $b_{lm}(\hat z)$ essentially resetting the
orientation of the beam (say $\rho(\hat n)=0$).

In this case, the orthogonality relation,
\begin{eqnarray}
\int^{2\pi}_0 d\phi\exp(-i (m_1+m_2) \phi) \ = \ 2\pi\,\delta_{m_1,-m_2}\,,
\end{eqnarray}
implies that the integral over $\phi$ in Eq.({\ref{eq:beam-BipoSH}}),
separates from the integral over $\theta$ and would restrict the
non-zero beam-BipoSH to $M=0$, 

\begin{eqnarray}\label{eq:BB}
B^{LM}_{l_1 l_2} = \delta_{M0}\frac{2\pi\prod_{l_1}}{\sqrt{4\pi}}\sum_{m'}b_{l_2 m'}(\hat z)\times \sum_{m_2}& (-1)^{m_2}C^{L0}_{l_1 -m_2 l_2 m_2}I^{l_1 l_2}_{m_2,m'}\,,
\end{eqnarray}

\noindent where

\begin{equation}\label{eq:integralI}
I^{l_1 l_2}_{m_2,m'} \ = \ \int d^{l_2}_{m_2 m'}(\theta)d^{l_1}_{m_2 0}(\theta)\sin\theta d\theta.
\end{equation}

\noindent Here we use the symmetry property of Wigner-\textit{d}
functions, $d^{l}_{m m'}=(-1)^{m-m'}d^{l}_{-m-m'}$.

To make analytical progress, we need to evaluate $I^{l_1
l_2}_{m_2,m'}$ for $m'=-2,0,2$ (as already discussed for all other $m'$ modes $b_{l_2m'}$ are negligable). 
 It is important to note that, circular part of beam function $m'=0$ will show up as $L=0$ mode in beam-BipoSH coefficient. The non-circular $m'=\pm2$ part of the beam
will give rise to non-trivial ($L\ne2$) beam-BipoSH.
\begin{itemize}
\item {\em beam-BipoSH due to $m'=0$ mode of beam function}:

For $m'=0$, $I^{l_1 l_2}_{m_2,m'}$ in Eq.(\ref{eq:BB}) can be simplified to,
\begin{eqnarray}\label{eq:integralI1}
I^{l_1 l_2}_{m_2,0}&=&\frac{2}{2l_2+1}\delta_{l_1 l_2}.
\end{eqnarray}
Therefore, beam-BipoSH coefficients for circular part of the beam are of following form (refer
Appendix~\ref{app:beam-biposh}),
\begin{eqnarray}\label{eq:beam-biposh-C}
B^{LM}_{l_1 l_2}&=&(-1)^{l_2}b_{l_2 0}(\hat z)\sqrt{4\pi}\delta_{l_1
l_2}\delta_{L0}\delta_{M0}\delta_{m'0}.
\end{eqnarray}
\item {\em beam-BipoSH due to $m'=\pm2$ mode of beam function}:

For $m'=\pm2$, the integrals are evaluated separately for the $m_2=0$
and $m_2\neq 0$ parts of the summation.
In the former case when $m_{2}=0$ and $m'=\pm2$, the integral in Eq.(\ref{eq:integralI}) simplifies to,
\begin{eqnarray}\label{eq:integralI2}
I^{l_1 l_2}_{0,\pm2}\ = \ 
\quad\left\{
\begin{array}{ll}
0 & \mbox{if ($l_1 +l_2\equiv \textrm{odd}$)} \\ 
0 & \mbox{if ($l_1 >l_2$)} \\ 
4\sqrt{\frac{(l_{2}-2)!}{(l_{2}+2)!}} & \mbox{if ($l_1 <l_2$)} \\ 
\sqrt{\frac{(l_{2}-2)!}{(l_{2}+2)!}}\big[\frac{4l_2}{(2l_2 +1)}-\frac{2l_2(l_2 +1)}{(2l_2 +1)}\big] & \mbox{if ($l_1 = l_2$)} \,.
\end{array}
\right.
\end{eqnarray}
For $m_2\neq0$, $d^{l_2}_{m_2
\pm 2}(\theta)$ is recursively expanded in terms of $d^{l_2}_{m_2
0}(\theta)$ to evaluate $I^{l_1 l_2}_{m_2,\pm2}$ (refer Appendix
\ref{app:beam-biposh}).
NC beam with reflection symmetry have non-vanishing beam-BipoSH with even-parity. 
Hence, the beam-BipoSH due to the NC part of the beam
in the PT-scan approximation is of the following form,
only,
\begin{eqnarray}
B^{LM}_{l_1
l_2}=\delta_{M0}\frac{2\pi\prod_{l_1}}{\sqrt{4\pi}}\left(b_{l_2 2}(\hat z)+b^{*}_{l_2 2}(\hat z) \right)  
\left[C^{L0}_{l_1 0 l_2 0}I^{l_1
l_2}_{0,2}+\sum_{m_2\neq0}(-1)^{m_2}C^{L0}_{l_1 -m_2 l_2 m_2}I^{l_1 l_2}_{m_2,2}
\right]\,.
\label{beamBiposhm2}
\end{eqnarray} 
For a PT scan, beam-BipoSH encodes the effect of NC beam $b_{l_2 2}(\hat z)$ in the second part of the expression.
\end{itemize}

The above expression for beam-BipoSH coefficient
holds for the PT-scan (with constant $\rho({\hat n})$)
for a NC-beam that has reflection symmetry. 
Although we have restricted explicit analytic results presented in the
text to reflection symmetric beam functions, in general, odd parity
beam BipoSH will be non-vanishing in absence of the above mentioned
symmetries. Appendix~\ref{app:beam-biposh} provides expressions for
odd-Parity beam-BipoSH $B^{LM^{(-)}}_{l_1 l_2}$, that can be used as a
measure of breakdown of reflection symmetry in NC
beam\footnote{Departure from reflection symmetry in the beam in a
full-sky CMB experiment, if ignored, also causes leakage of power from
the $\sim 500$ times stronger CMB dipole signal into higher multipole,
most importantly, contaminating the CMB quadrupole moment of the
angular power spectrum. This has been studied and estimates on WMAP
beam maps indicates the effect of reflection breakdown symmetry is
expected to be small, but not negligible~\cite{SD-TS2012a}.}. Note that
the BipoSH estimator~\cite{DH-AL}, that differ by a factor from
original definition of Hajian \& Souradeep~\cite{AH-TS-03,AH-TS-06},
used by the WMAP team cannot be extended to odd-parity BipoSH,
However, it is possible to devise BipoSH estimators that can measure
odd-parity BipoSH spectra while matching that employed by WMAP for
even-parity BipoSH spectra~\cite{MK-TS2010}.

\section{ Relating  ${\rm beam-BipoSH}$ with ${\rm CMB-BipoSH}$}\label{bipforbeam}

The measured CMB temperature is the convolution of true underlying CMB temperature 
with the instrument beam,

\begin{eqnarray}
\widetilde{\Delta T}(\hat n_{1})=\int d{\Omega_{\hat n_{2}}}B(\hat
n_{1},\hat n_{2})\Delta T(\hat n_{2}).
\end{eqnarray}
Here, $\Delta T(\hat n_2)$ is the underling true sky temperature along $\hat{n}_2$ and $\widetilde{\Delta T}(\hat n_1)$
is the temperature measured along $\hat{n}_1$. 
  $B(\hat n_{1},\hat n_{2})$ is known as the beam response function and gives the
sensitivity of the detector around the pointing direction, $\hat n_{1}$.  The
observed two point correlation function is,
\begin{eqnarray}
\tilde C(\hat n_{1},\hat n_{2})\equiv\langle\widetilde{\Delta T}(\hat
n_{1})\widetilde{\Delta T}(\hat n_{2}) \rangle \ =\int d\Omega_{n}\int
d\Omega_{n'}C(\hat n',\hat n) B(\hat n_{1},\hat n') B(\hat n_{2},\hat
n),
\label{eq:obs-corr}
\end{eqnarray}
where $C(\hat n',\hat n)=\langle\Delta T(\hat n')\Delta T(\hat
n)\rangle$, is the underlying correlation function. 
It is evident from Eq.({\ref{eq:obs-corr}}), that SI
violation can occur either due to breakdown of rotational invariance of the
underlying correlation function $C(\hat n_{1},\hat
n_{2})\not\equiv C(\hat n_{1}\cdot \hat n_{2})$, or due to the
breakdown of circularity in beam response function $B(\hat n_{1},\hat
n_{2})\not\equiv B(\hat n_{1}\cdot \hat n_{2})$, or both.

Inverse transform of Eq.(\ref{eq:BPOSH}), yields the most general for CMB-BipoSH
coefficients $\tilde A^{LM}_{l_1 l_2}$,

\begin{eqnarray}
 \tilde A^{LM}_{l_1 l_2}&=&\int d\Omega_{n_1}\int d\Omega_{n_2}\tilde
 C(\hat n_{1},\hat n_{2})\{Y_{l_{1}}(\hat{n}_{1})\otimes
 Y_{l_{2}}(\hat{n}_{2})\}^{*}_{L M} 
\label{biposhbeam}
\end{eqnarray}

\noindent which under PT scan of an underlying isotropic sky  becomes (Appendix~\ref{app:cmb-biposh})

\begin{eqnarray}
\tilde A^{LM}_{l_1 l_2} \ =\ \sum_{l}(-1)^{l} C_{l} \sum_{L_1 M_1 L_2 M_2}B^{L_1
M_1}_{l_1 l}B^{L_2 M_2}_{l_2 l} \ \times 
\prod_{L_1 L_2}C^{LM}_{L_1 M_1
L_2 M_2} {\begin{Bmatrix} l & l_2 & L_2 \\ L & L_1 & l_1
\end{Bmatrix}}\,.
\label{biposhbeamM0}
\end{eqnarray}

The equations shows that provided the beam-BipoSH coefficients ($B^{LM}_{ll'}$) are restricted to $M=0$, 
the corresponding BipoSH coefficients of the CMB maps are also restricted to $M=0$. 
It also turns out that due to triangularity condition ($|L_1-L_2|\leq L\leq L_1+L_2$),
the most dominant terms in the above summation are $\{L_1=L,
L_2=0\}\ \textrm{and}\ \{L_1=0, L_2=L\}$ as they are proportional to $B^{0 0}_{l_1 l}B^{L 0}_{l_2 l}$ and 
$B^{L 0}_{l_1 l}B^{0 0}_{l_2 l}$. The product of these two beam-BipoSH coefficients in turn depends on the product of SH coefficients $b_{l0}b_{l2}$.
In a mildly non-circular beam response function $b_{l0}$, is significantly larger than $b_{l2}$,
 making $b_{l0}b_{l2}$ much larger than $b_{l2}b_{l2}$, which will 
contribute as second order terms in 
Eq.(\ref{biposhbeamM0}).

 The BipoSH estimator used by the WMAP team~\cite{CB-RH-GH,DH-AL},
differs from our defination (the original BipoSH definition in Hajian \&
Souradeep~\cite{AH-TS-03}) by a factor of ${\prod_{L}}/({\prod_{l_1
l_2}C^{L0}_{l_1 0 l_2 0})}$ and are restricted to only even-parity
BipoSH~\footnote{Note that this factor in WMAP-BipoSH estimator
strictly restricts BipoSH considerations to the even parity sector
since $C^{L0}_{l0 l'0} =0$ for odd values of the sum $L=l+l'$. In the
context of NC-beams, this would be a handicap if reflection symmetry
is violated leading to odd-parity BipoSH coefficients. Also it is blind to a number of other
interesting possibilities with odd-BipoSH signals. }

\begin{eqnarray}\label{biposh-constantscan}
\tilde A^{L0^{\rm WMAP}}_{l_1 l_2} = \frac{\prod_{L}}{\prod_{l_1
l_2}C^{L0}_{l_1 0 l_2 0}} \tilde A^{L0}_{l_1 l_2}\,.
\label{Bmatrix}
\end{eqnarray}

SI violation signals in WMAP-7 were measured in two BipoSH spectra,
$A^{20}_{ll}$ and $A^{20}_{l-2 l}$, we provide explicit leading order
expressions for these coefficients arising from the NC-beam as,
\begin{eqnarray}
\label{eq:WMAP1}
\tilde A^{20^{\rm WMAP}}_{ll} & = &\frac{(-1)^{l}\,2\sqrt{5}\,C_{l}\,B^{00}_{ll}B^{20}_{ll}}{(\prod_l)^{3} C^{20}_{l
0 l 0}}\, ,\\
\tilde A^{20^{\rm WMAP}}_{l-2 l} &=&\frac{\sqrt{5}(-1)^{l}}{\prod_{{l-2}
l}C^{20}_{{l-2} 0 l 0}} \ \times 
\  \Big[\frac{C_{l-2}B^{00}_{{l-2} {l-2}}B^{20}_{l
{l-2}}}{\prod_{l-2}}+ \frac{C_{l}B^{00}_{l l}B^{20}_{{l-2}
l}}{\prod_{l}} \Big]\,.\label{eq:WMAP2}
\end{eqnarray}

Note, that BipoSH expression in Eq.(\ref{eq:WMAP1}) and
Eq.(\ref{eq:WMAP2}) are provided in the scaled form that matches the
BipoSH estimator employed by the WMAP team. The BipoSH spectra plotted in different figures in this paper are in accordance with WMAP definition.

\section{Analytical evalution of ${\rm CMB-BipoSH}$ coefficients}\label{scanning}

\subsection{Elliptical Gaussian beam in PT scan}
\label{app:EG}

Elliptical-Gaussian (EG) functions provide a simple model of NC-beam as an
extension to the often used circular-symmetric Gaussian beam
function. 
BipoSH coefficients obtained from EG beams serve to crosscheck and validate 
analytical expression derived in Eq.(\ref{eq:BB}), Eq.(\ref{eq:WMAP1}) and  Eq.(\ref{eq:WMAP2}), and puts a check on the numerical simulation of
CMB maps convolved (in real space) with an NC-beam (full details of numerical simulations can be found in \cite{Das2014,SD-TS2012a}).

An EG-beam function
pointed along $\hat z$ axis can be expressed in spherical polar
coordinates, as
\begin{equation} 
B(\hat{z},\hat{n}) \ = \ \frac{1}{2\pi\sigma_1\sigma_2}\exp \left[ -\frac{{\theta^2}}{2\sigma^2(\phi)} \right],
\end{equation}
where the azimuth angle dependent beam-width $\sigma(\phi_1) \ \equiv
\ [\sigma_1^2/(1+\epsilon \sin^2\phi_1)]^{1/2}$ is given by Gaussian
widths $\sigma_1$ and $\sigma_2$ along the semi-major and semi-minor
axes. The non-circularity parameter $\epsilon \ = \
(\sigma_1^2/\sigma_2^2 - 1)$, which  is related to eccentricity $e = \sqrt{1-\sigma_2^2/\sigma_1^2}$. 
As expected, the EG beam reduces to circular Gaussian
beam for zero eccentricity ($e=0$). Higher the value of eccentricity,
stronger the deviation from circularity. 
An analytical expression for the beam-SH of EG-beam  is available in~\cite{TS-BR}. Due to  even-fold azimuthal symmetry and 
reflection symmetry, $b_{lm}(\hat z) = 0$, for odd $m$;  and for even $m$,
\begin{eqnarray}\label{eq:blm-EG}
&&b_{lm}(\hat z)\ \ = \ \ \sqrt{\frac{(2l+1)(l+m)!}{4\pi(l-m)!}}(l+\frac{1}{2})^{-m} \ \times \\
&& \quad I_{m/2}\Bigg[
(l+\frac{1}{2})^{2}\frac{\sigma^{2}_{1}e^{2}}{4}\Bigg]\exp\big[{-(l+\frac{1}{2})^{2}\frac{
\sigma^{2}_{1}}{2}({1-\frac{e^{2}}{2}})\big]} \, , \nonumber
\end{eqnarray}
where $I_\nu(x)$ is the modified Bessel function. 
The reality condition of beam,
$b^{*}_{lm}=b_{lm}$ for even $m$, then implies $b_{l -m}=b_{lm}$.
\begin{figure}[h]
\centering
\includegraphics[height=0.5\textwidth,angle=-90]{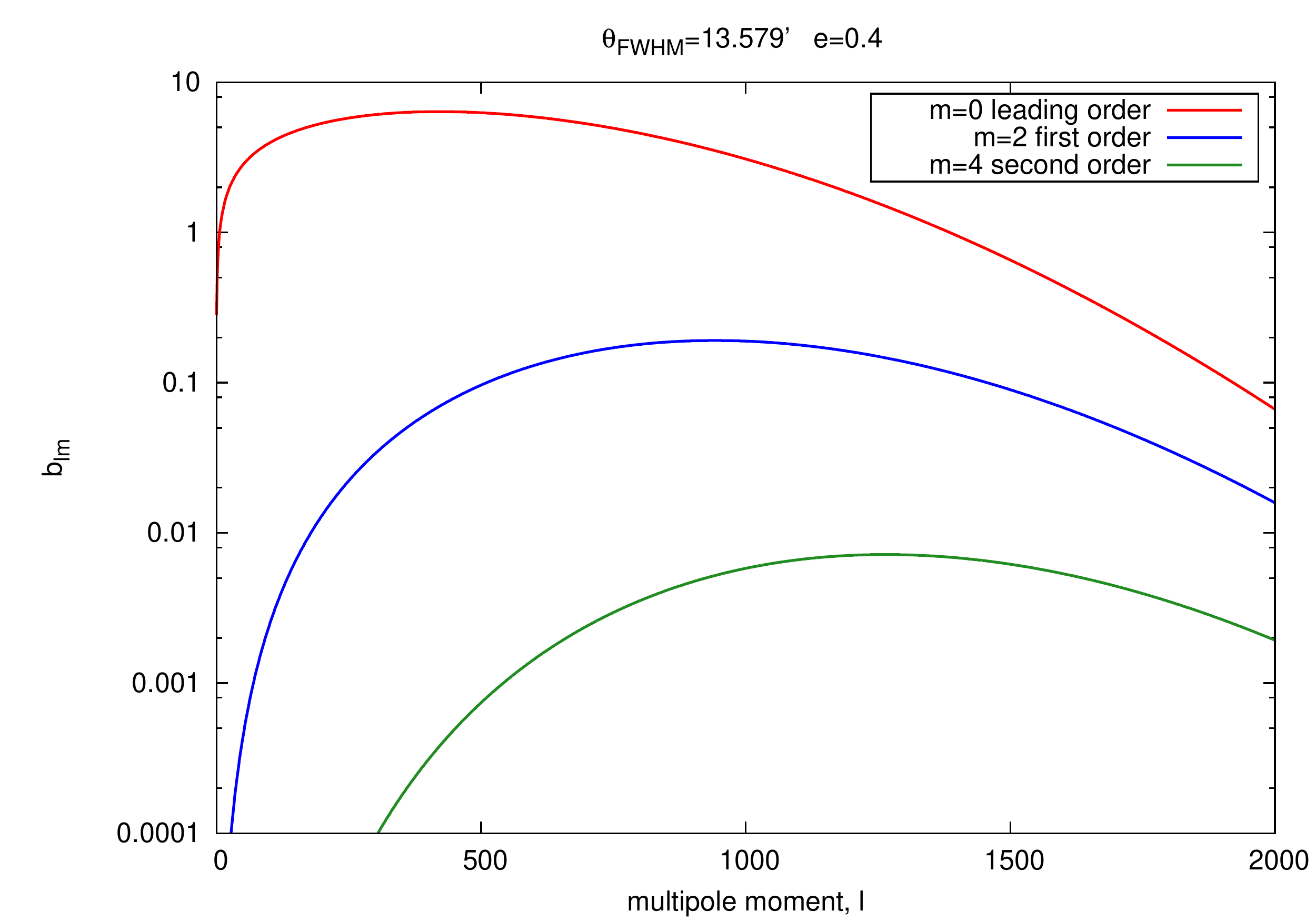}
\caption{The coefficients of spherical harmonic expansion of an EG beam with
$\theta_{\rm{FWHM}}=13.579'$ and eccentricity $e=0.4$. The circular
symmetric component of the beam, given by $m=0$ is the dominating term. The next leading 
contribution comes from  $m=2$ mode which gives the effect of non-circularity of the beam. The higher 
$m$ modes are negligible.}
\label{fig:ellgauss-blm}
\end{figure}

\begin{figure}
\centering
\includegraphics[height=0.45\textwidth, angle=-90,trim = 0 0 0 10, clip]{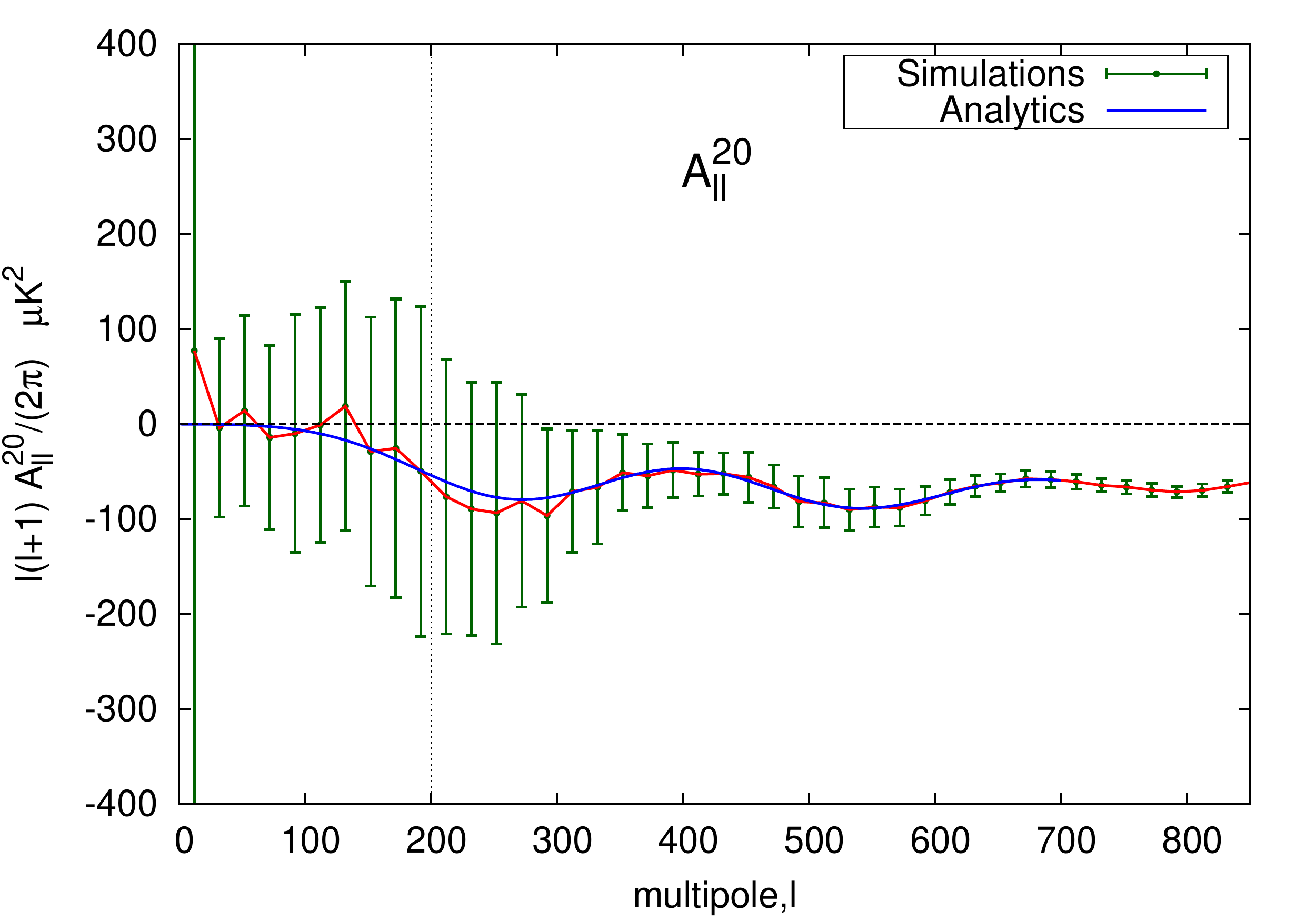}
\includegraphics[height=0.45\textwidth,angle=-90,trim = 0 0 0 10, clip]{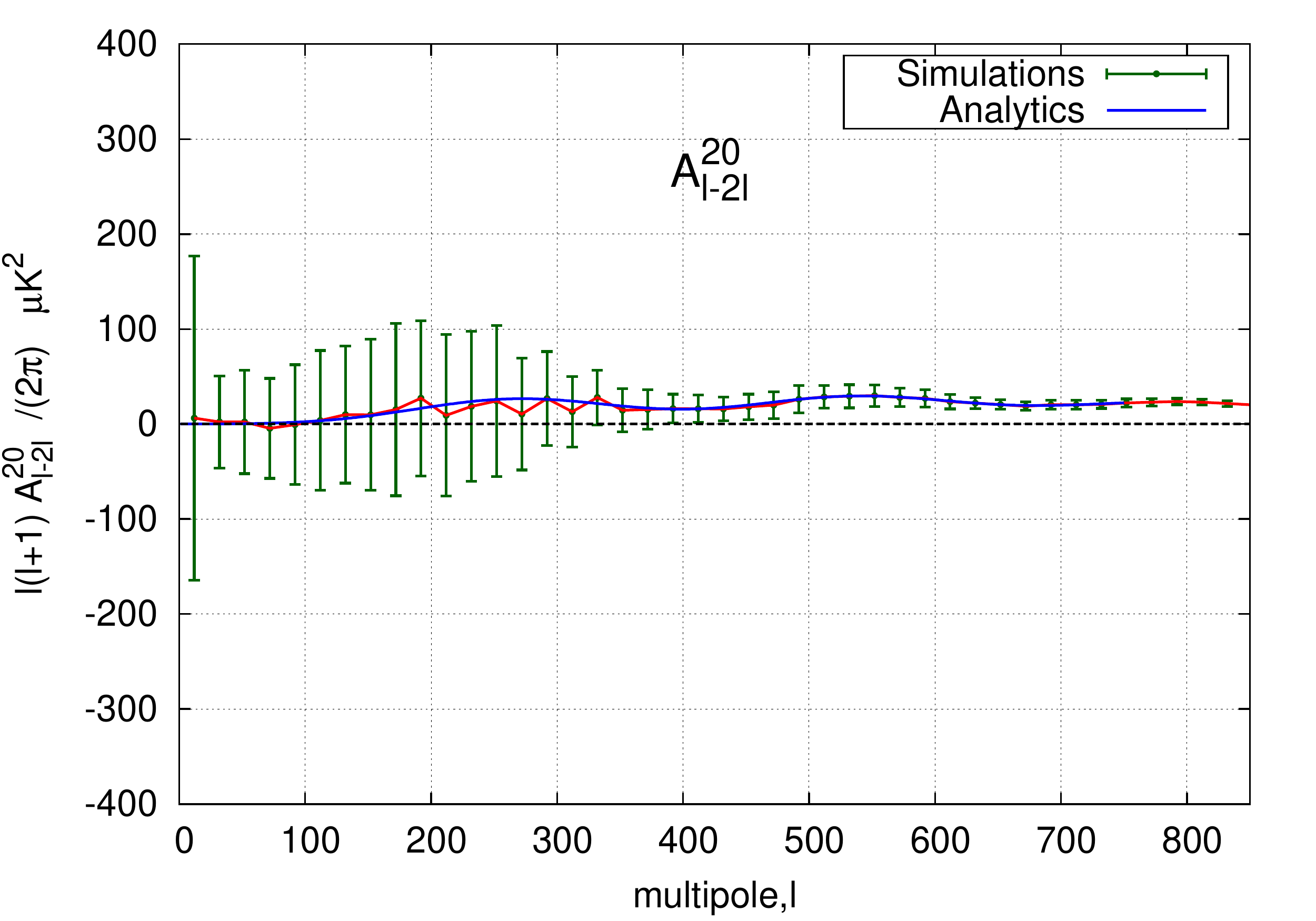}

\includegraphics[height=0.45\textwidth,angle=-90,trim = 0 0 0 10, clip]{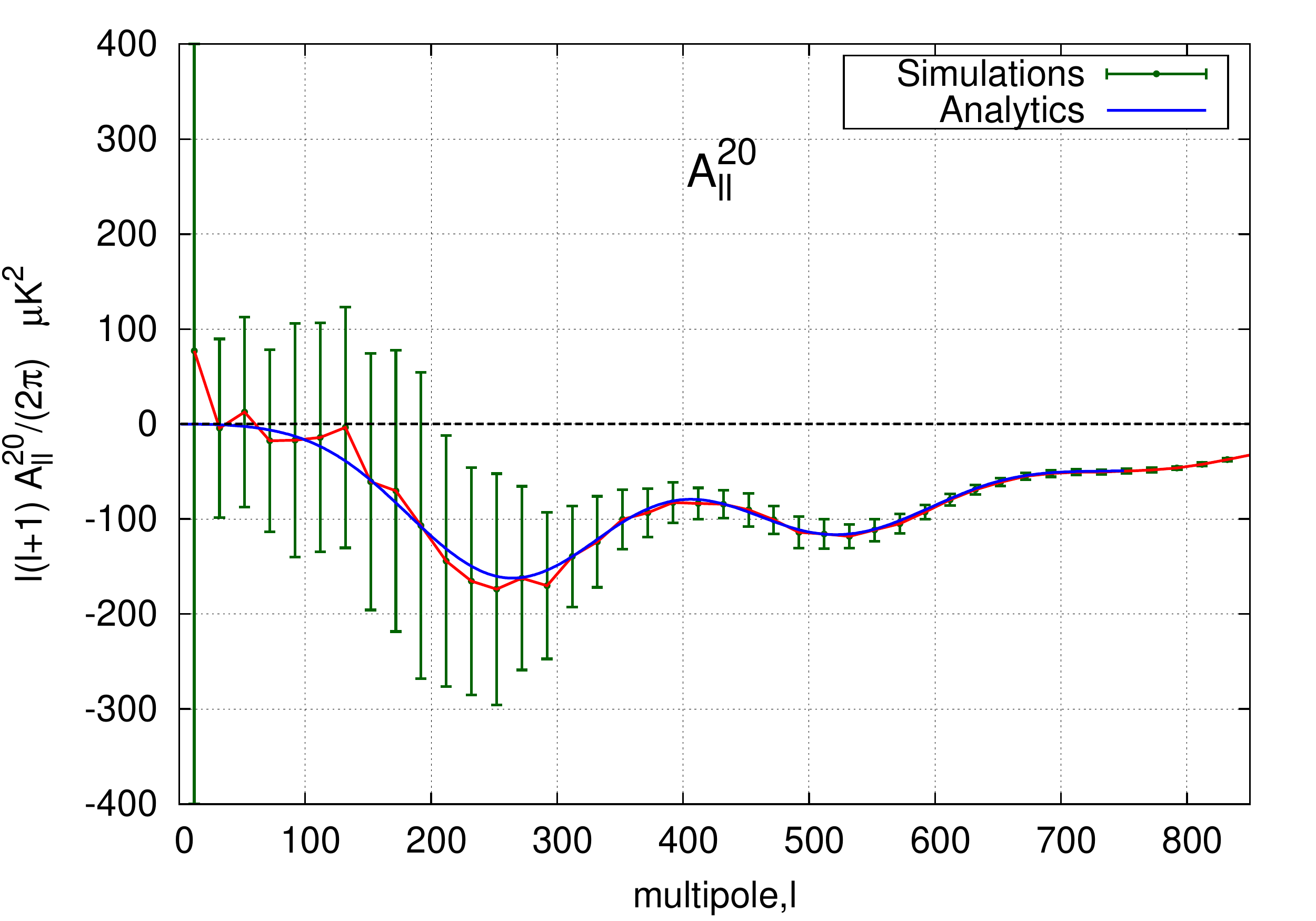}
\includegraphics[height=0.45\textwidth,angle=-90,trim = 0 0 0 10, clip]{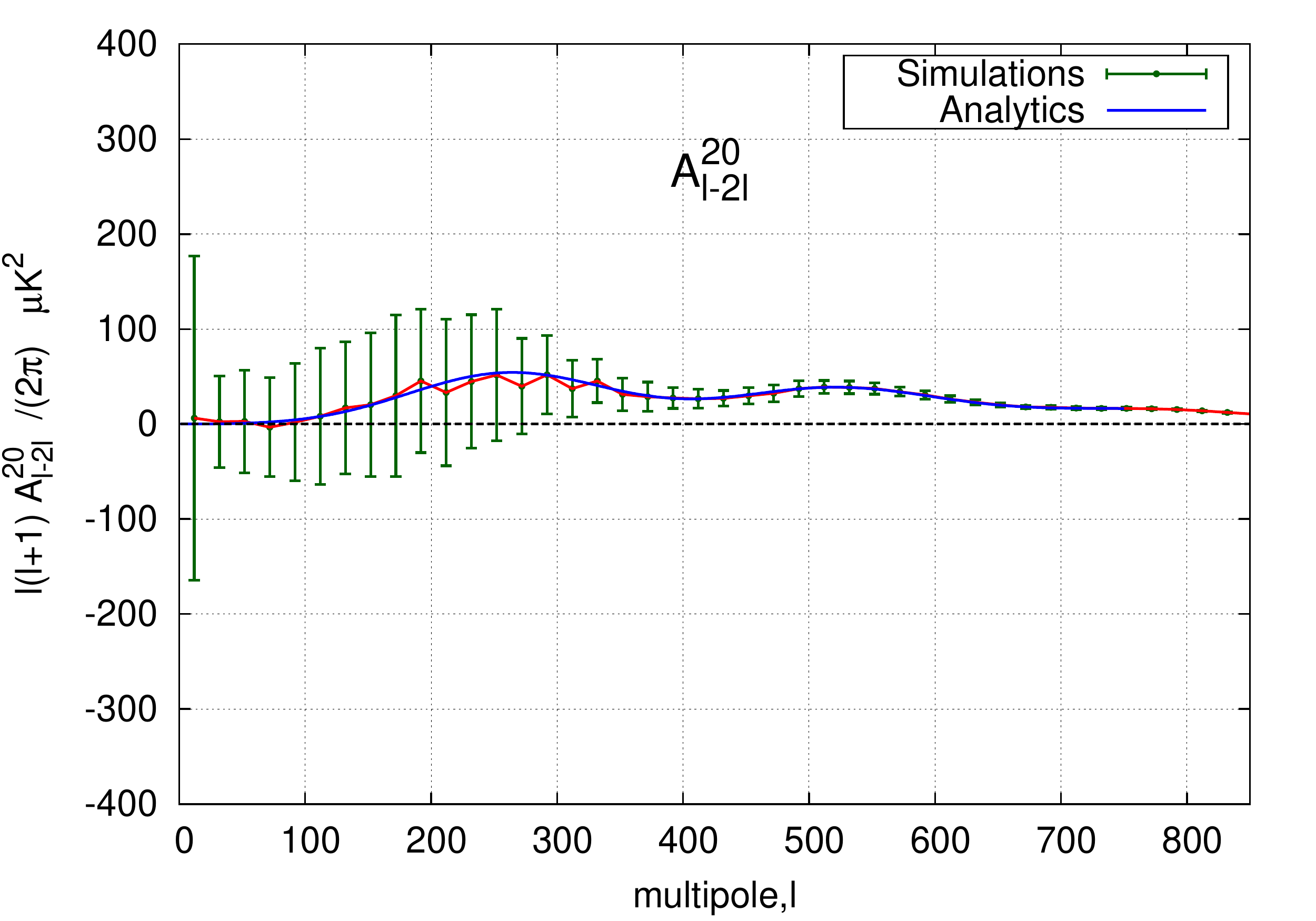}
	\caption{BipoSH spectra $A^{20}_{ll}$ and $A^{20}_{l-2 l}$ for
	 EG beam with : ({\em Top})
	 $\theta_{\rm{FWHM}}=13.579'$ and eccentricity $e=0.4$;
	 ({\em Bottom}) $\theta_{\rm{FWHM}}=17.7036'$ and
	 eccentricity $e=0.46$. Smooth (blue) curve are 
	 CMB-BipoSH computed using the analytic expressions. The red
	 curve with corresponding error-bars are obtained from 100
	 numerical simulated SI maps convolved with the same
	 EG beam. 
	 \label{fig:ellgauss-WV}}
\end{figure}

For EG beams, the ratio $b_{lm}/b_{l0}$ dies down rapidly with $|m|$. 
In Fig.~\ref{fig:ellgauss-blm}, we plot beam-SH coefficients of
an EG beam with $\theta_{\rm{FWHM}}=13.579'$ and eccentricity $e=0.4$, 
which is close to an elliptical estimate of 
W band beam of WMAP. The plot clearly shows that the $m=4$ mode  
is negligible in compared to $m=2$ mode. We will also get similar feature if we consider 
an EG-beam with $\theta_{\rm{FWHM}}=17.7036'$ with eccentricity $e=0.46$ which is close to the V-band beam. 
Therefore, for our analysis we restrict our calculations to $m=0,\pm2$ modes.  We estimate beam-BipoSH coefficient in
Eq.(\ref{eq:BB}) for PT scan by using the closed analytical form of $b_{lm}$'s as in Eq.(\ref{eq:blm-EG}).  Finally,
using Eq.(\ref{eq:WMAP1}) and Eq.(\ref{eq:WMAP2}), we obtain the CMB
BipoSH spectra $A^{20}_{ll}$ and $A^{20}_{l-2 l}$. We verify our analytical results 
with BipoSH coefficients evaluated from $100$ SI maps numerically convolved with  EG-beam functions 
(see  Fig.~\ref{fig:ellgauss-WV}).

\subsection{WMAP raw beam with PT scan}

\begin{figure*}
\centering
\includegraphics[width=0.45\textwidth,trim = 0 0 0 10, clip]{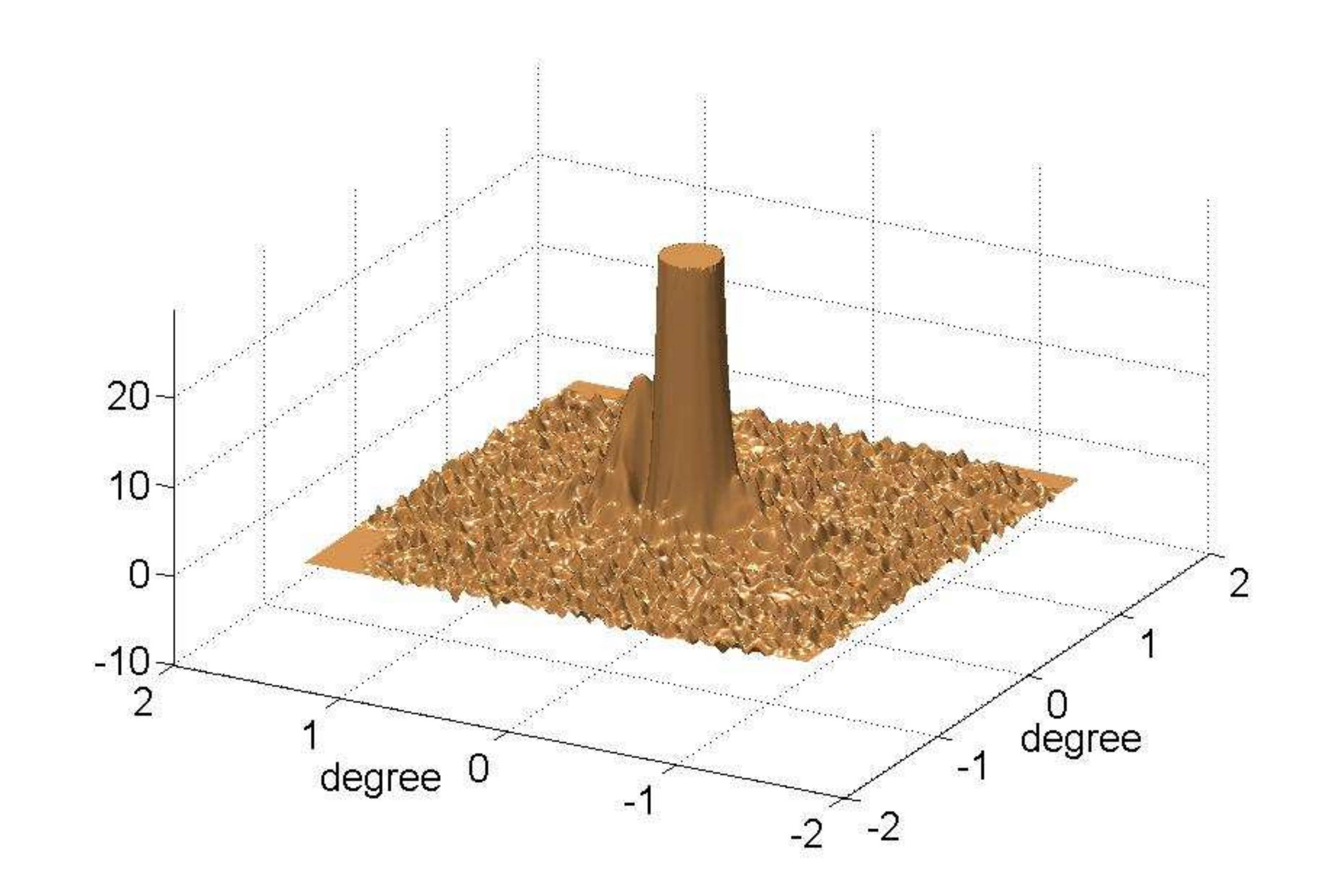}
\includegraphics[width=0.45\textwidth,trim = 0 0 0 10, clip]{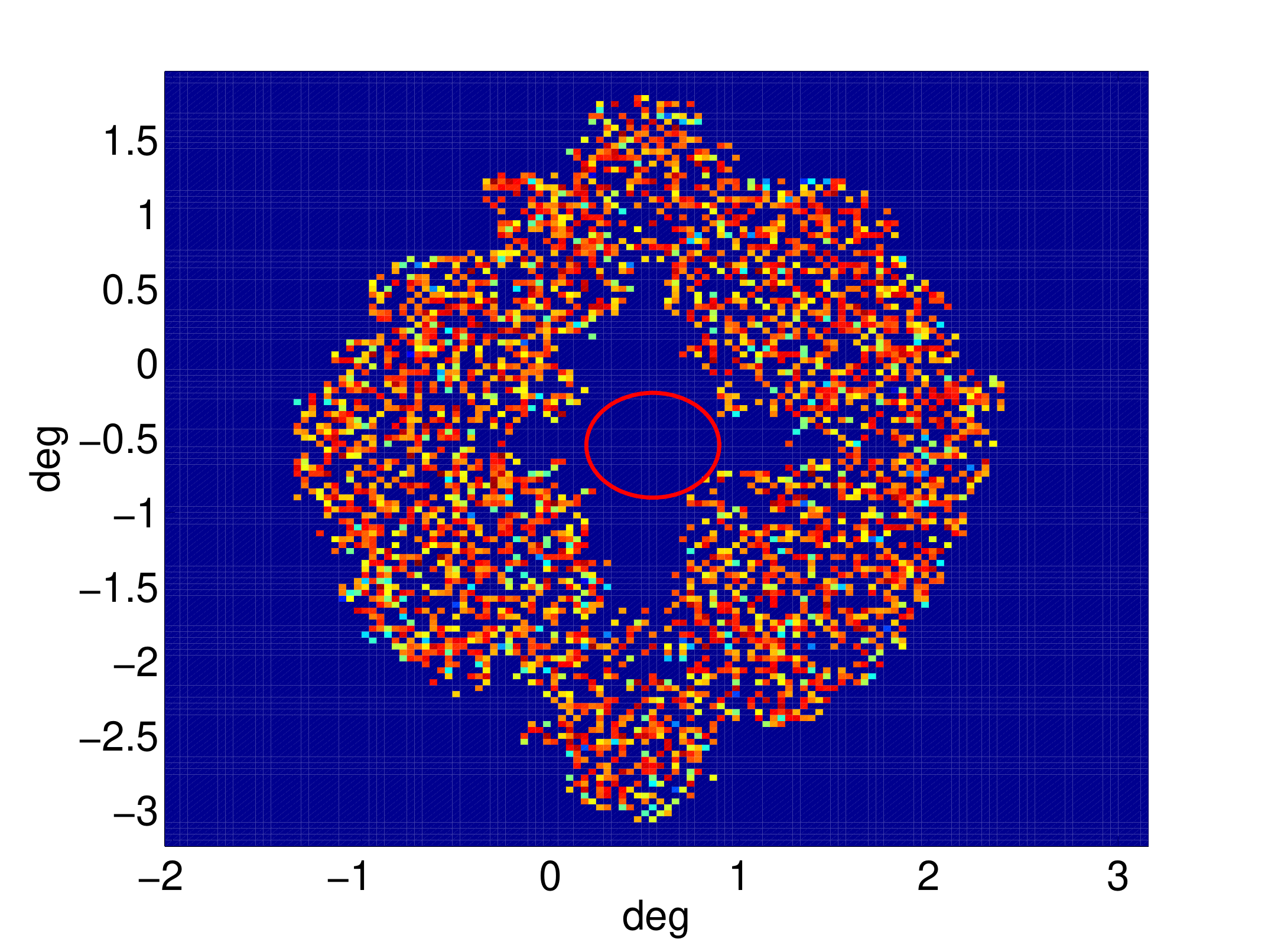}

\includegraphics[width=0.45\textwidth,trim = 0 0 0 10, clip]{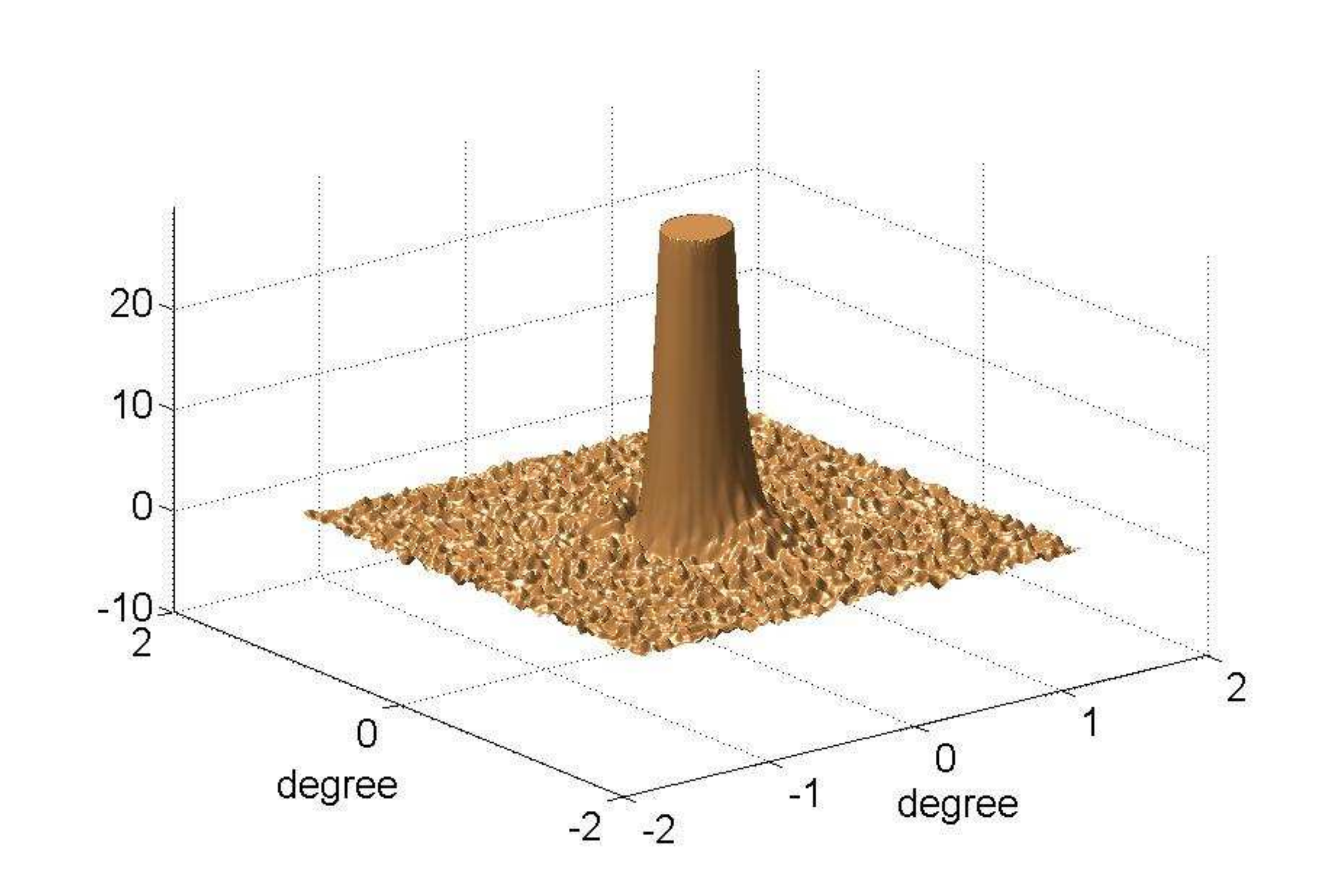}
\includegraphics[width=0.45\textwidth,trim = 0 0 0 10, clip]{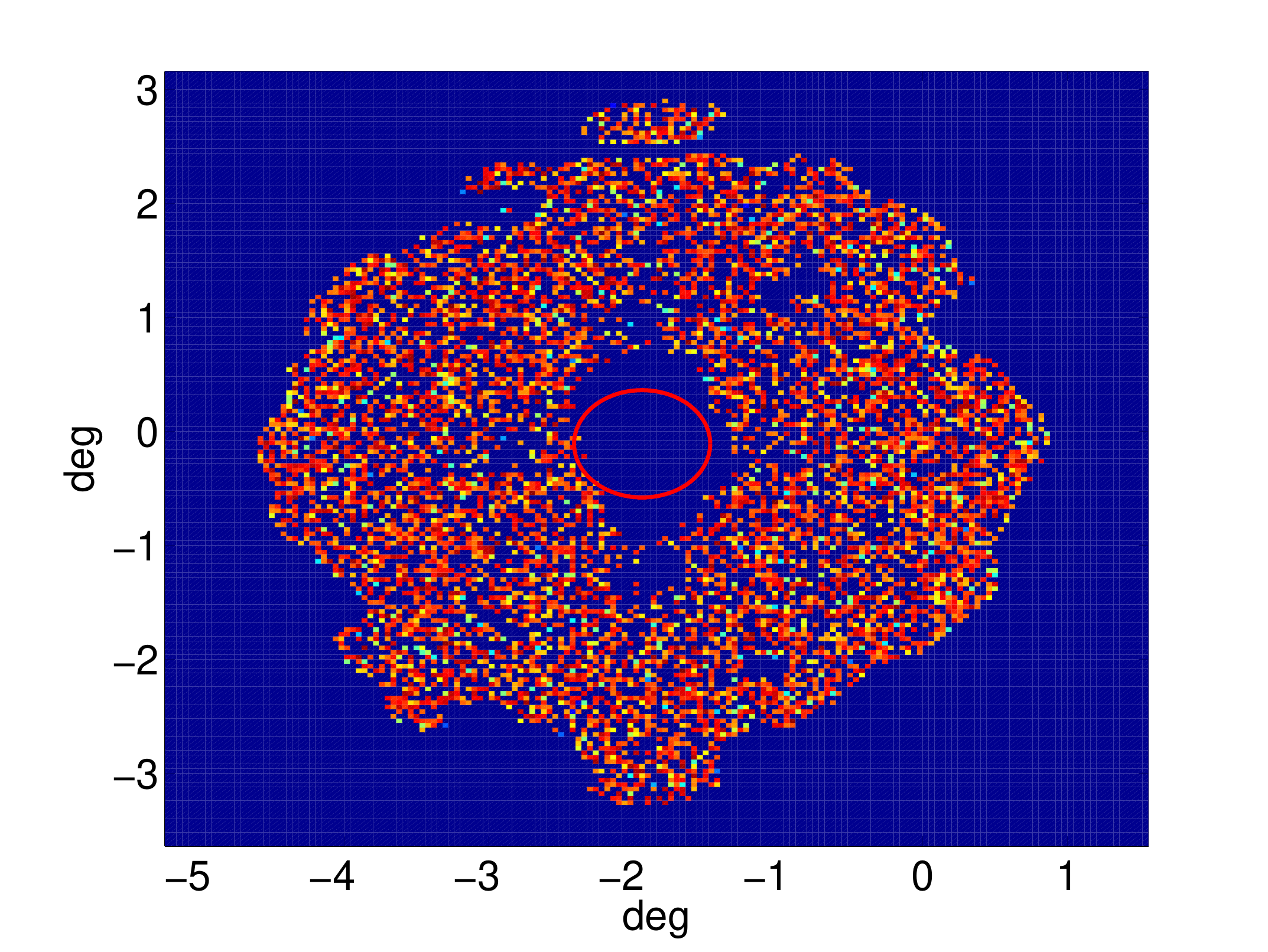}
	 \caption{Beam response function of A side of W1 differencing
assembly {\em (Top)} and A side of V2 differencing assembly {\em
(Bottom)} are shown. The {\em Left-hand} panels show 3D zoomed view of the central part of the beam. 
The central part is truncated to highlight the features existed in the elliptical contour. 
The {\em Right-hand} panels cover the entire beam-map images to show the
spread-out annular distribution of regions with negative response.
(The red circles marks the central beam peak region). The power in
negative response is $\sim 0.5$ of the positive power in the central
peak. The annular negative power distribution shows quadrupolar
feature that modifies the beam-SH $b_{l2}$ to take negative values at
high $l$ (see Fig~\ref{fig:W1-V2fullblm}). 
It is apparent then that correctly accounting
for WMAP NC-beam effects numerically, requires
the convolution of the entire beam map region with the SI
sky-map leading to enormous increase in computing costs (relative to
using only the central peak).}
\label{fig:BRA-W1V2}	 
\end{figure*}

\begin{figure*}[!htbp]
\centering
\includegraphics[width=0.9\textwidth]{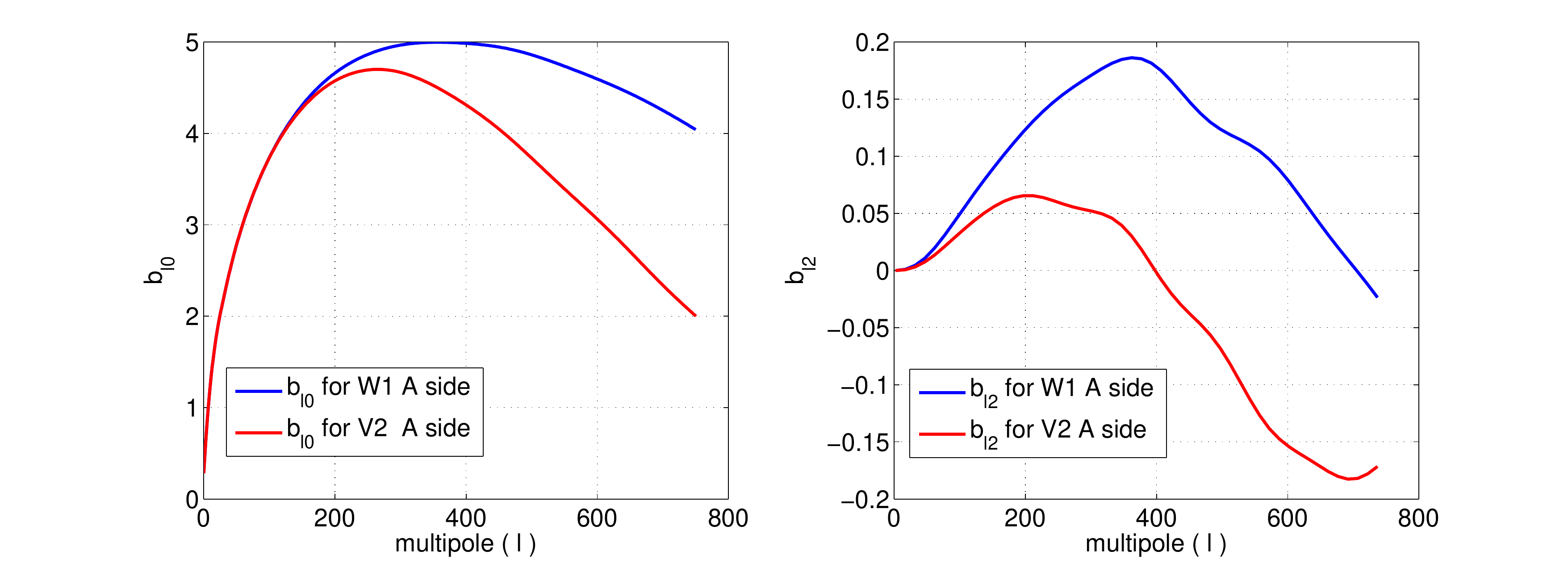}
\includegraphics[width=0.8\textwidth,trim = 0 260 0 260, clip]{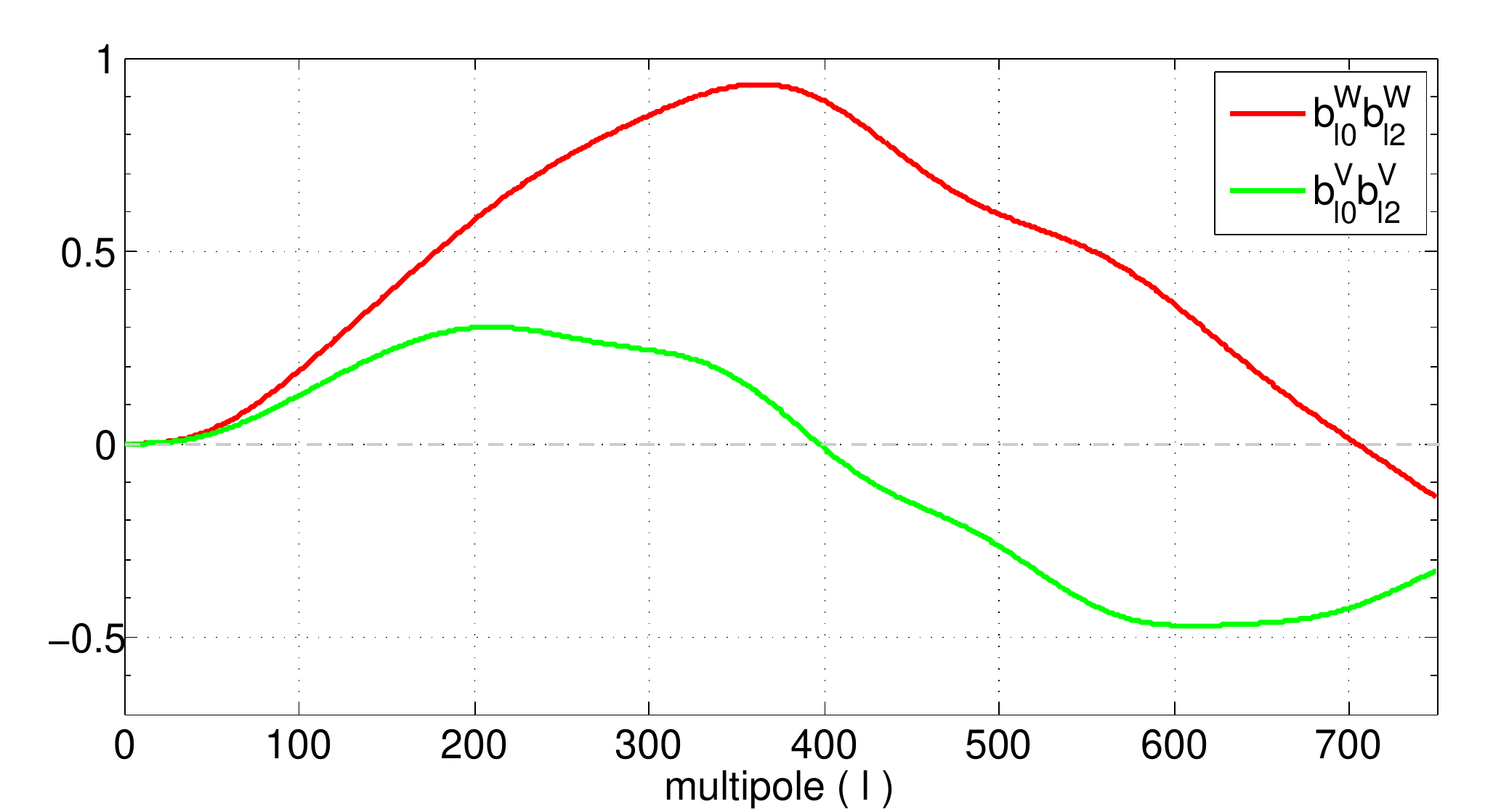}
\caption{ {\em Top:} Beam spherical harmonic transforms, $b_{l0}$ and
$b_{l2}$ of beam maps of A side of W1 and V2 DA. 
\newline {\em Bottom}: The plot of the NC-beam leading
order NC beam perturbation parameter $b_{l0} b_{l2}$ vs. $l$ . 
The plots show that although BipoSH peak structure is largely set by the underlying
angular power spectrum $C_l$ of SI cosmological mode, small differences observed at the two
different frequencies can arise because of the difference in the shapes of $b_{l0}b_{l2}$.}
\label{fig:blm-W1-V2}
\end{figure*}

\label{BipoSHWMAPrawbeam}

It is widely-known that the WMAP beams are non-circular and deviate from a Gaussian profile and
that an EG beam is not a good
approximation~\cite{Pag_WMAP03,SM-AS-TS,Hin_WMAP07}. 
WMAP-7 year data had a whopping SI violation detection in  $A^{20}_{ll}$ and $A^{20}_{ll-2}$ BipoSH coefficients~\cite{CB-RH-GH}. 
Later on it was realized that it was due to the noncircularity of the WMAP beams which was corrected in the WMAP-9 year data. 
No such signal is observed in Planck data which reinforce the fact that it was due to the particular shape of the beam and the scan pattern.

To see the imprint of the WMAP kind of beam on the BipoSH coefficients,  we consider the A side raw beam maps
of the V2 and W1 differencing Assembly (DA) of WMAP as representative
of the V and W band beams, respectively (see Fig.~\ref{fig:BRA-W1V2}). 
The central part of the beam maps show an elliptical peak with 
non-trivial `shoulder-like' features. Apart from this, the beam functions contain an annular region 
with positive and negative sensitivity spread over a diameter of $3^{\circ}$ to $5^{\circ}$. 
In the right-hand panels, we highlight the 
regions with negative response. The integrated
power in the negative beam response is $\sim0.5$ of the total power.

We compute the beam-SH
coefficients for these assemblies numerically to use in semi-analytic estimate
of the CMB-BipoSH coefficients, using Eq.(\ref{eq:WMAP1}) and
Eq.(\ref{eq:WMAP2}). 
Fig.~\ref{fig:blm-W1-V2} is a plot of the
$b_{l0}$ and leading order $b_{l2}$ beam-SH coefficients of
the W1A and V2A raw beam maps.  Note that the {\it $b_{l2}$ spectrum changes
sign and takes negative values at high $l$} -- a key qualitative
feature of the WMAP beams.  The origin of this curious feature is the negative responce of the beam sensitivity 
function as seen in the right hand panel of the Fig.~\ref{fig:BRA-W1V2}. 
This particular feature cannot be captured in
Elliptical-Gaussian beam model where $b_{l2}$ does not change sign with $l$.
This peculiar nature of $b_{l2}$ reflects as  flipping of sign in $A^{20}_{ll}$ spectra of 
V-band at high $l$ as seen in the WMAP-7 measurements. 
Such a unique correspondence between a beam-SH
feature and the consequent CMB-BipoSH is unlikely to be mimicked by
other effects and has been confirmed independently by various authors that 
WMAP seven year SI violation detection was due to NC Beam.

\begin{figure*}[h]
\centering
\includegraphics[height=0.45\textwidth, angle=-90,trim = 0 0 0 10, clip]{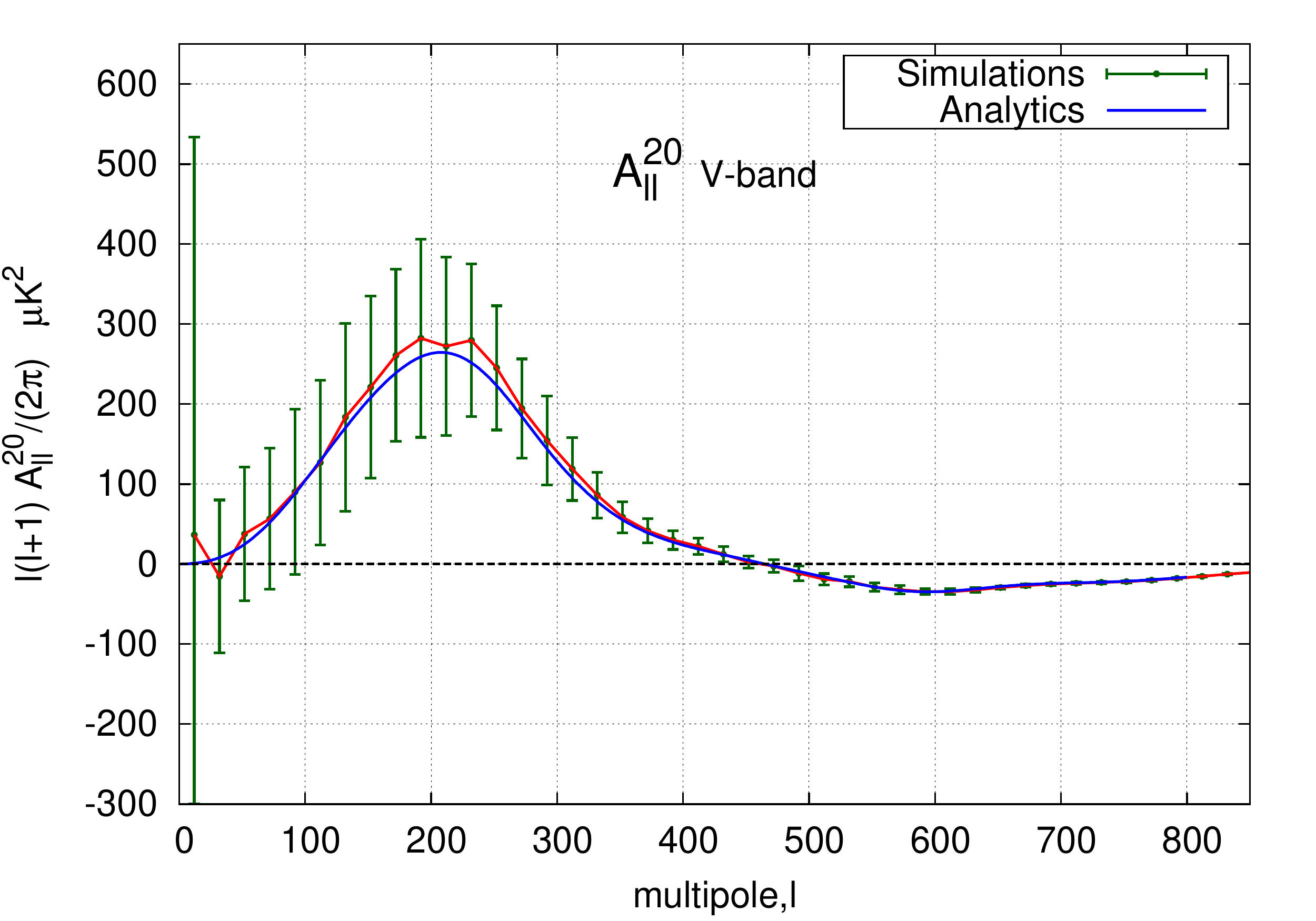}
\includegraphics[height=0.45\textwidth, angle=-90,trim = 0 0 0 10, clip]{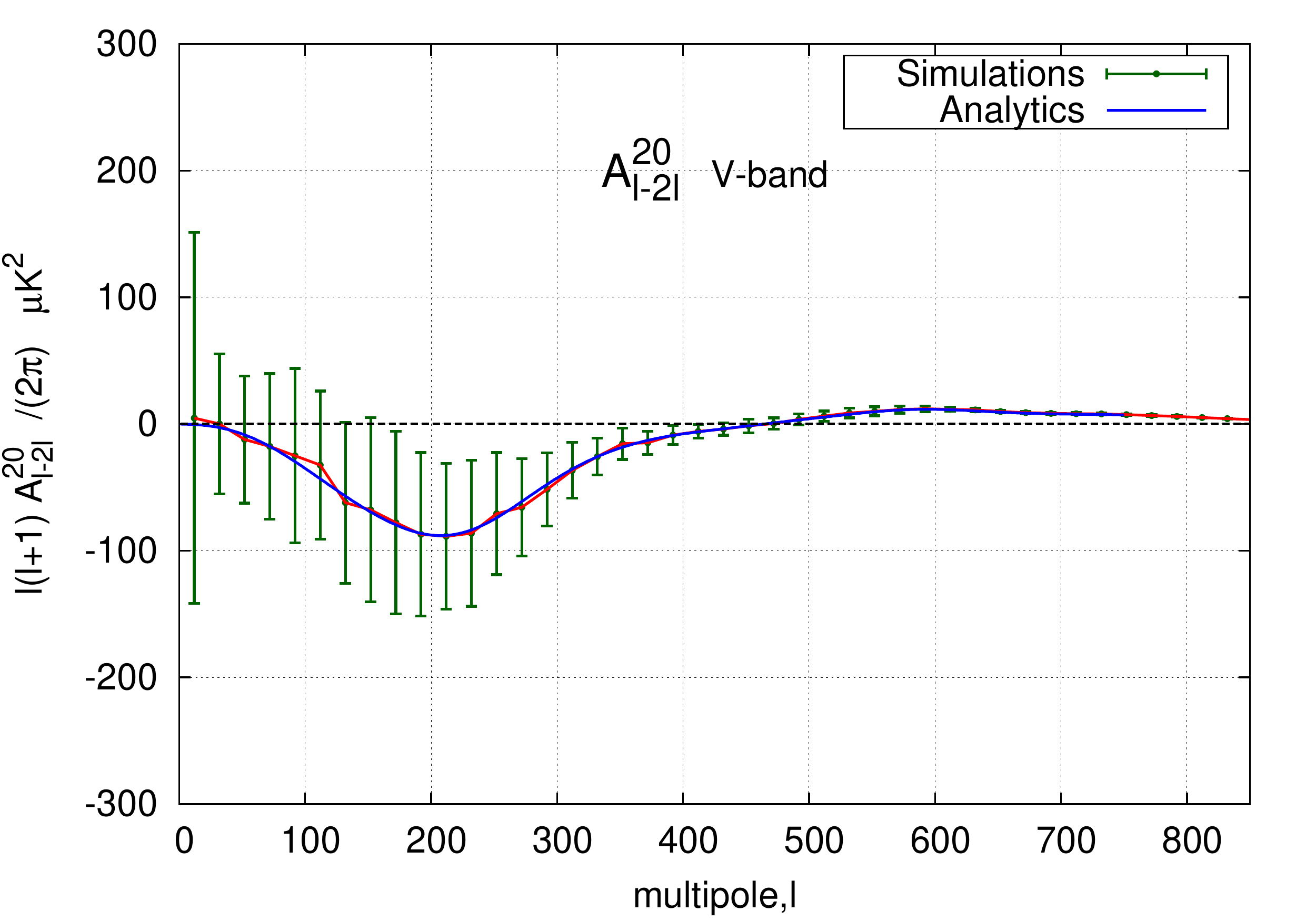}

\includegraphics[height=0.45\textwidth, angle=-90,trim = 0 0 0 10, clip]{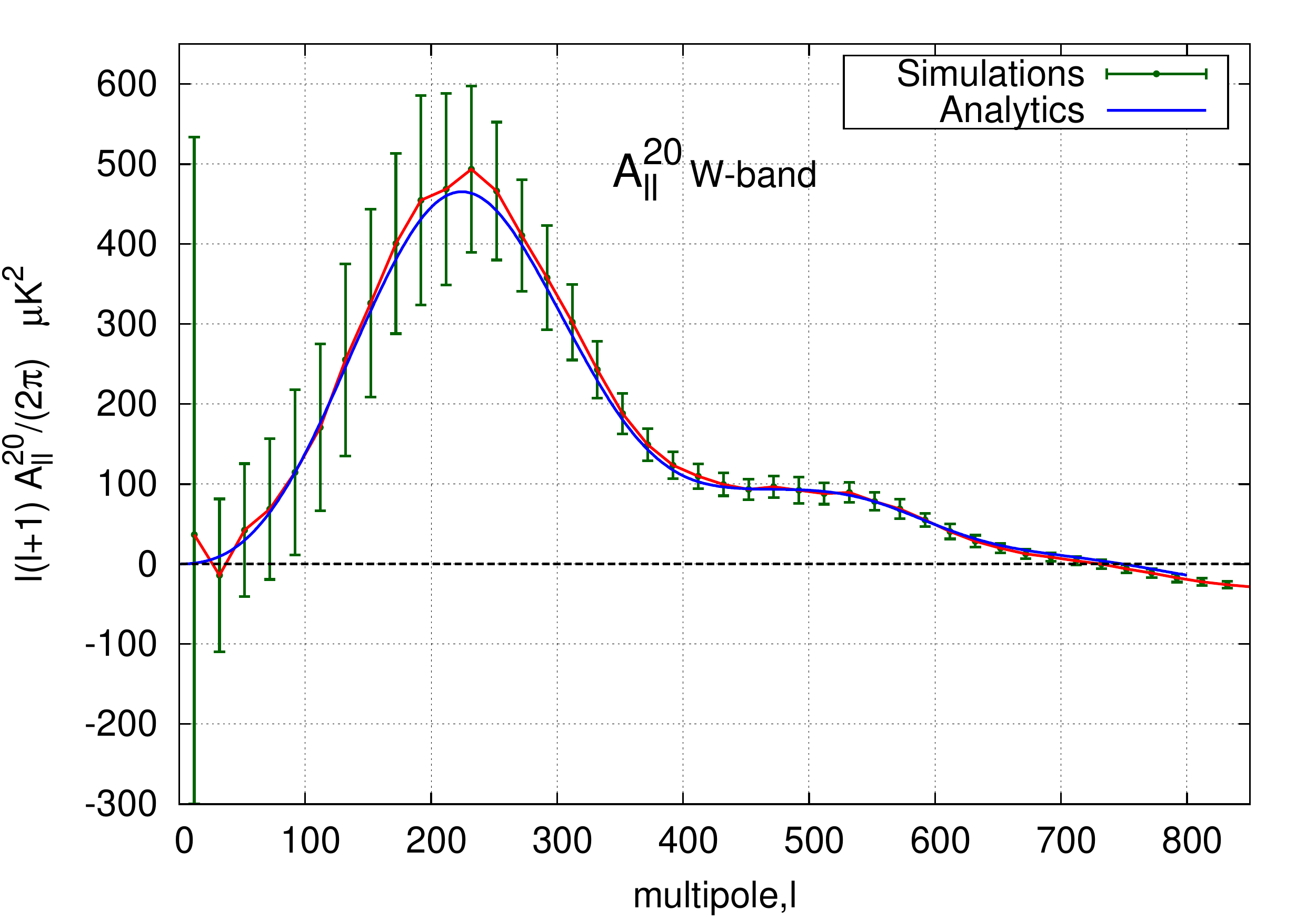}
\includegraphics[height=0.45\textwidth, angle=-90,trim = 0 0 0 10, clip]{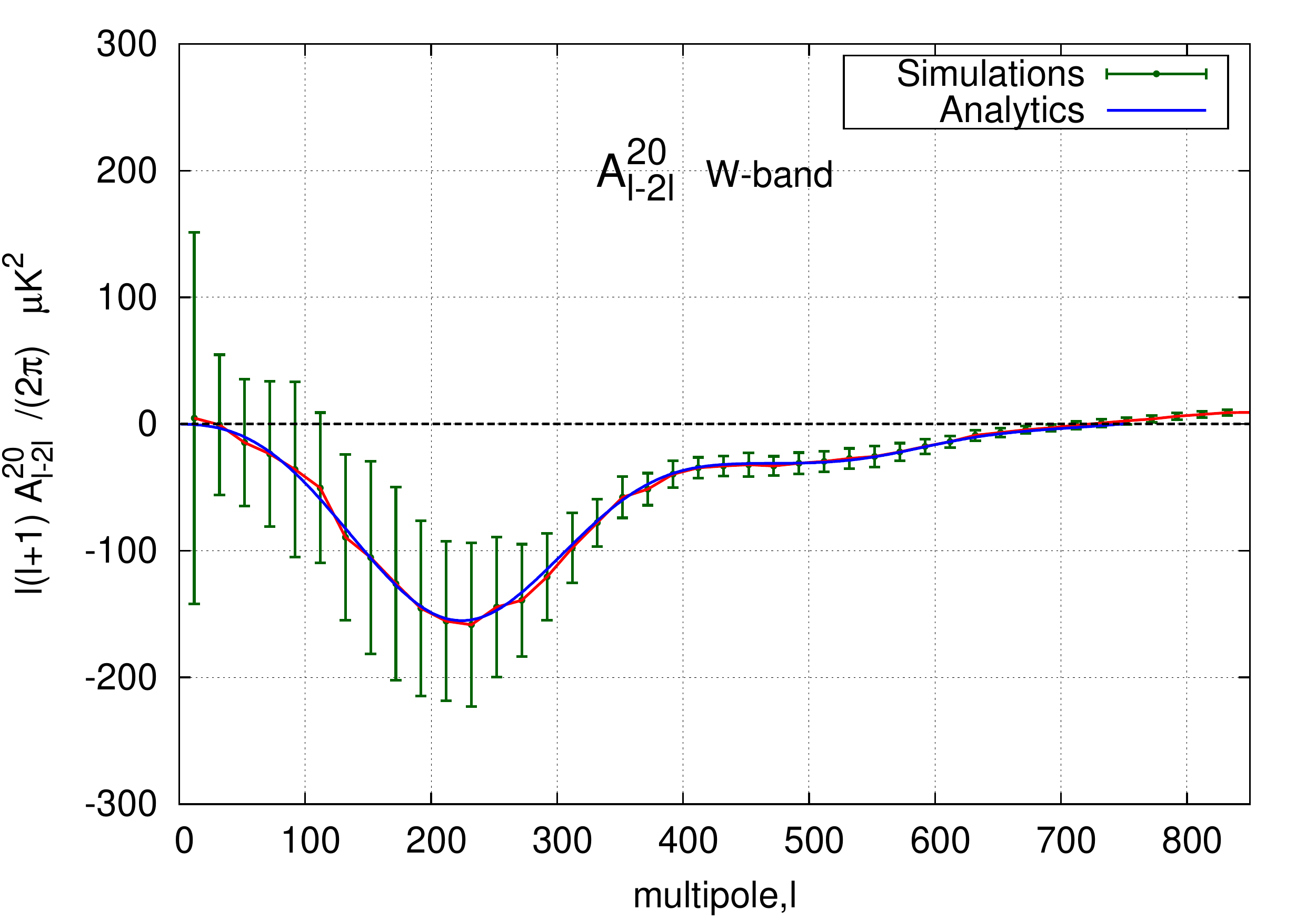}
\caption{BipoSH spectra, $A^{20}_{ll}$ ({\em Left}) and $A^{20}_{l-2
l}$ ({\em Right}) obtained for the raw beam maps A side of V2
channel ({\em Top}) and A side of W1 channel ({\em Bottom}) of the
WMAP experiment.  Analytically evaluated BipoSH spectra (blue)
overlaid on the average BipoSH spectra (red) with error bars obtained
from 100 simulations of statistically isotropic CMB sky convolved W1
and V2 channel of WMAP. }
\label{fig:W1-V2fullblm}
\end{figure*}

We use PT-scan  in ecliptic coordinates to get an estimate of $A^{20}_{ll}$ and $A^{20}_{ll-2}$ due to NC Beam.
This particular coordinate system is chosen because WMAP scan is azimuthally symmetric around ecliptic pole. 
We verify our analytical results using BipoSH coefficients from the numerical simulations where we generate non-SI maps 
by numerically convolving the WMAP W1A and V2A beam with the SI maps (see Fig.~\ref{fig:W1-V2fullblm}).

Here we list some of the key features that we see in the results :

\begin{enumerate}
\item{} From our analytical understanding, which is also verified by numerical simulations, in a coordinate system where PT scan is valid, only $M=0$
should be significant. 
\item{} We notice the NC beam effect is larger in W band than in V
band explaining the difference in detected SI violation signal at the
two frequencies.
\item{} The BipoSH spectra $A^{20}_{ll}$, $A^{20}_{ll-2}$  
changes sign at large $l$ (in V-band its more prominent). This happens because the beams 
contains the negative sensitivity region in the annular part leading to a sign flip in $b_{l2}$ at high $l$.
\item{} The BipoSH coefficients from NC beam shows a prominent bump
roughly around the first acoustic peak ($l=220$) for both W and V
band. This corresponds mainly to the scale picked by the underlying
angular power spectrum $C_l$. However, the precise peak location also
depends on the peak in $b_{l0} b_{l2}$ for each band and can account
for differences in the peak location in the two bands shown in
Fig.~\ref{fig:blm-W1-V2}.
\end{enumerate}


\subsection{WMAP raw beam and scan strategy}
\label{Sec:WMAPapp}
The WMAP satellite follows a differential scan strategy where it  records the temperature difference between 
two telescopic horns for each frequency band. 
The pair of horns are about $70.5^{\circ}$ off the satellite's symmetry axis. It spins 
with a spin period of around 2.2 minutes about the symmetric axis. Along with this the spacecraft also has 
a slow precession, $22.5^{\circ}$ about the Sun-WMAP line. Precession period is about 1 hour. 
The satellite orbits around the sun with a period of a year~\cite{SD-TS2012a}. 

The differential scan pattern is mostly used to reduce the noise in the observed data. However, for evaluating an analytical estimate of the BipoSH coefficients for the real scan, we make the following assumptions : 
\begin{itemize}
\item{}We assume the beams in the A side and B side of the differential assembly are identical.  
\item{}The observed temperature of a pixel is the average of all the hits on that pixel from any orientation of the beam ( irrespective of A or B side ). 
\end{itemize}

The effect of non-circular beam on sky map is sensitive to the scan-strategy.
The effective beam that convolves the true sky resulting in the observed sky map is created by an intricate combination of the instantaneous beam of the instrument and the scan strategy.
Every time the beam hits a particular pixel, the beam BipoSH coefficient is given by Eq.(\ref{eq:beam-BipoSH}). 
The beam visits the same pointing direction ($\hat n'$) multiple times with different orientations $\rho_{{\hat n'},j}$, where $\rho_{{\hat n'},j}$ is the orientation of the beam at the $j^{\rm th}$ hit. The Beam BipoSH coefficients for this case can be expressed as  

\begin{eqnarray}
B^{LM}_{l_1 l_2}(\hat n',j,\hat z) =\sum_{m'}b_{l_2
m'}(\hat z)\Big(\sum_{m_1 m_2}C^{LM}_{l_1 m_1 l_2 m_2}
\int
\exp^{-im_2\phi}
d^{l_2}_{m_2
m'}(\theta)\exp^{-im'\rho_{\hat n',j}}Y^{*}_{l_1
m_1}(\hat n)~ d\hat n~\Big)\,. 
 \label{eq:genbeambiposh}
\end{eqnarray}

\noindent Here $\hat z$ denotes the direction along which the beam is expanded in the spherical harmonics and the beam-BipoSH 
coefficients are calculated. In case of PT scan, as  the beam visits different pixels with same orientation all the time, $\rho_{{\hat n'},j}$ = constant, which makes the BipoSH coefficient independent of the direction (${\hat n'}$), which is not true in general, as seen in the above equation. 
 The orientation of the beam for different directions ($\hat n'$) at different time ($j$) will be different.
Therefore, the beam-BipoSH becomes a function of the direction $\hat n'$ and time of scan $j$. 
As $\rho_{\hat n',j}$ is independent of $\hat n$ it comes out of the integral.

Suppose the beam hits the $\hat n'$ direction, $n_{\hat n'}$ times. As we assume that the temperatures along $\hat n$ direction are averaged over all the measurements, we can consider the beam along $\hat n'$ direction as an average of  beams in all the hits. Then the average beam-BipoSH can be expressed as 

\begin{eqnarray}
\bar B^{LM}_{l_1 l_2}(\hat n',\hat z) &=&\sum_{m'} \chi_{m'}(\hat n') b_{l_2 m'}(\hat z)\Big(\sum_{m_1 m_2}C^{LM}_{l_1 m_1 l_2 m_2}\times
\int \exp^{-im_2\phi} d^{l_2}_{m_2 m'}(\theta)Y^{*}_{l_1
m_1}(\hat n)~ d\hat n~\Big) \nonumber \\
&=&  \sum_{m'} \left(\sum_{lm}f^{(m')}_{lm}Y_{lm}(\hat n')\right){\mathcal B}^{LM^{({m'})}}_{l_1 l_2}(\hat z) \,,
 \label{eq:genbeambiposh}
\end{eqnarray}

\noindent where $\chi_{m'}(\hat n')  = \frac{1}{{n_{\hat n'}}}{\sum_j\left(\exp^{-im'\rho_{\hat n',j}}\right)}$.
Since for each $m'$ mode, $\chi_{m'}(\hat n')$ is a function of $\hat n'$, we expand them in the spherical harmonics (see 
Eq.(\ref{eq:genbeambiposh})). 
${\mathcal B}^{LM^{({m'})}}_{l_1 l_2}(\hat z)$ would be the beam-BipoSH coefficients if only one $m'$ mode of the beam was non zero with PT scan.

The $\hat n'$ direction of the sky is now scanned by a beam with beam-BipoSH coefficient $\bar B^{LM}_{l_1 l_2}(\hat n',\hat z)$. 
Therefore, substituting Eq.(\ref{eq:genbeambiposh}) into Eq.(\ref{Tn1}) the coefficients of the spherical harmonics of the scanned sky-map turns out to be

\begin{equation}
\tilde{a}_{JK}=\sum_{l_{1}m_{1}lmLM\bar{l}\bar{m}}\left(-1\right)^{m}\frac{\Pi_{\bar{l}}\Pi_{l_{1}}}{\sqrt{4\pi}\Pi_{j}}\sum_{m'}{\mathcal B}_{l_{1}l}^{LM^{(m')}}f_{\bar{l}\bar{m}}^{(m')}C_{l_{1}m_{1}l-m}^{LM}C_{\bar{l}0l_{1}0}^{J0}C_{\bar{l}\bar{m}l_{1}m_{1}}^{JK}a_{lm}\,.
\label{eq:ajk}
\end{equation}

\begin{figure}
\centering
\includegraphics[width=0.48\textwidth]{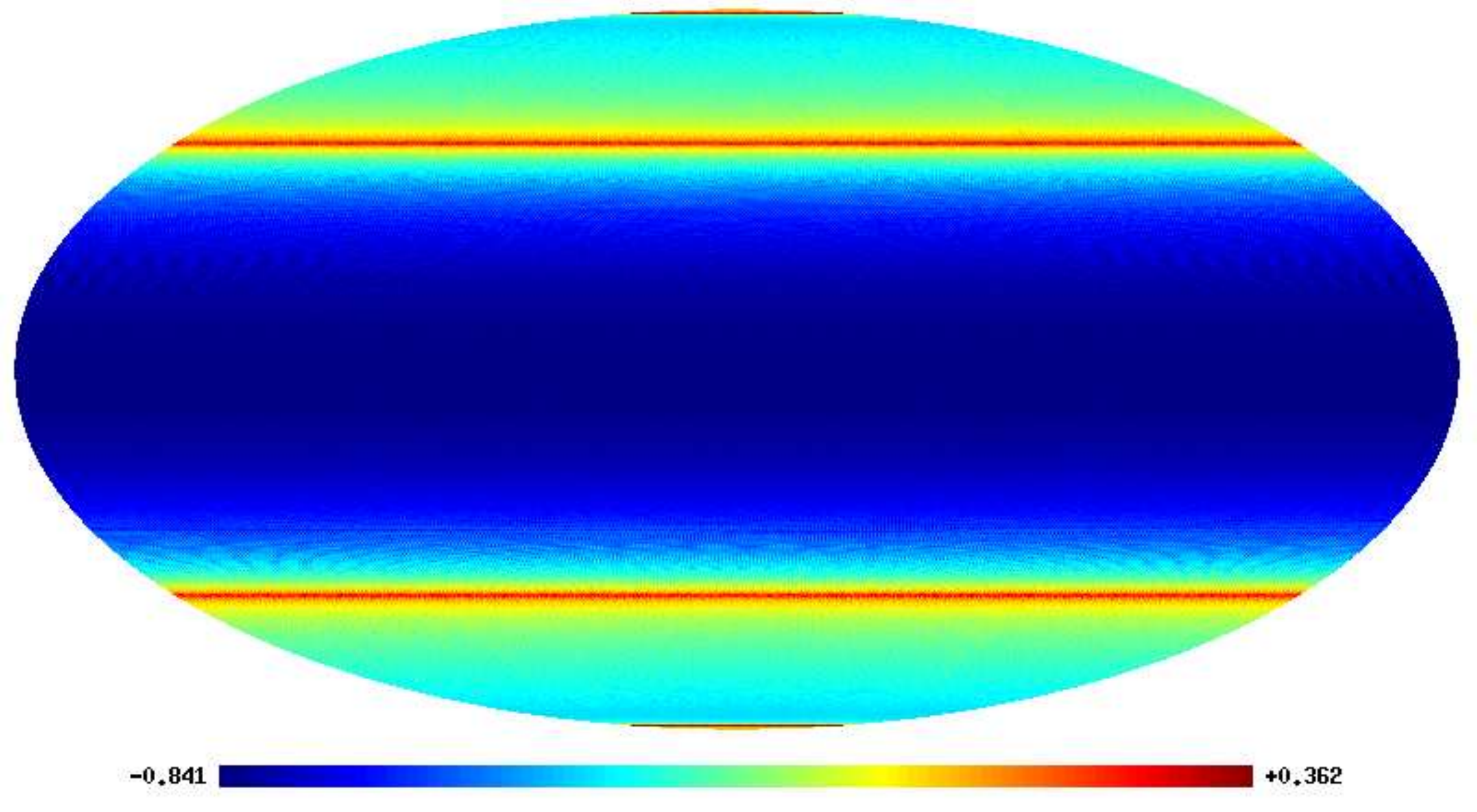} 
\includegraphics[width=0.48\textwidth]{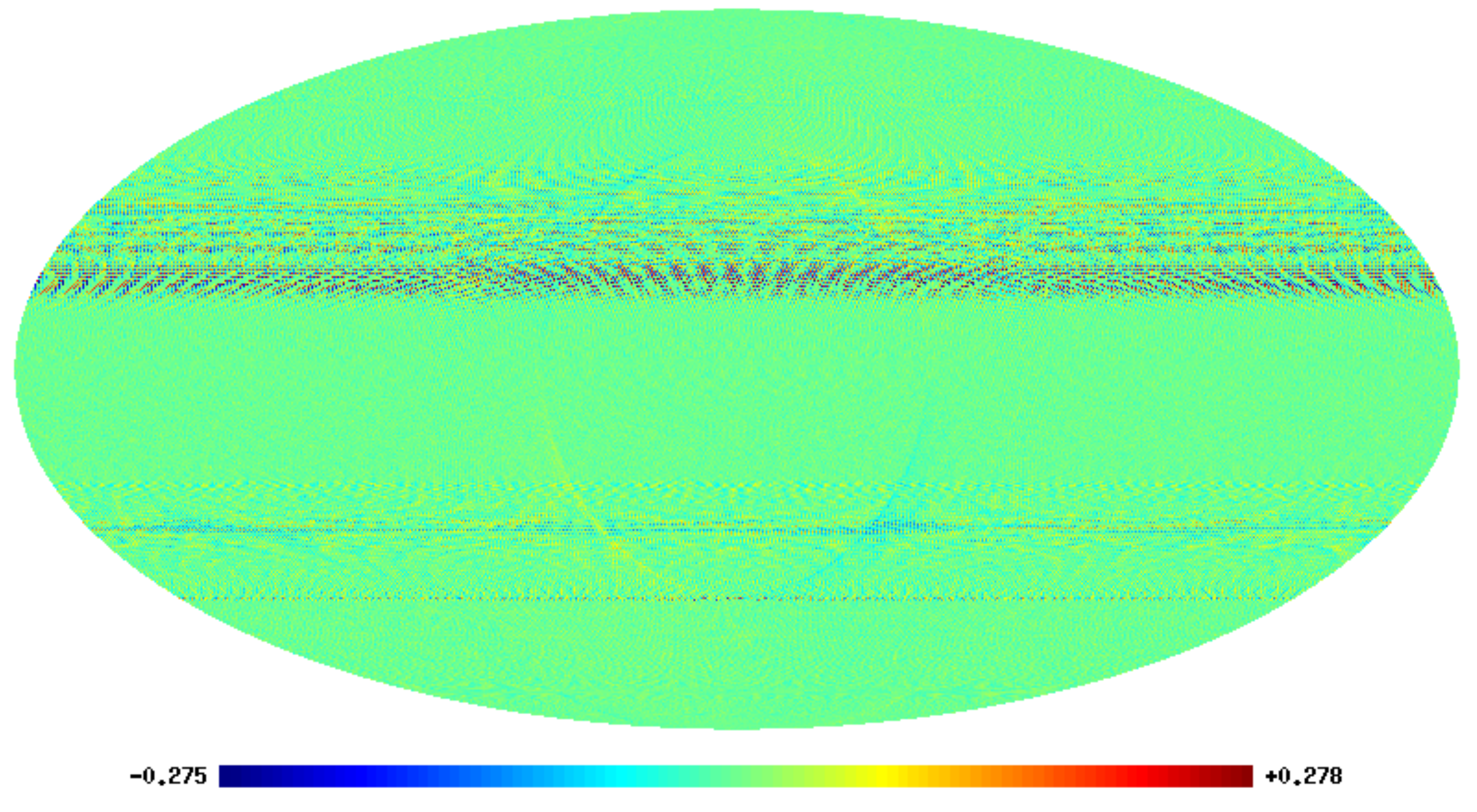} 
\caption{
Map of $\langle \cos(2\rho) \rangle$ and $\langle \sin(2\rho) \rangle$ in ecliptic
coordinate for full year WMAP scan. The angular bracket ( $\langle \,\dots\, \rangle$ ) denotes 
average over all the hits on a pixel irrespective of beam from A-side or B-side differential 
assambly of WMAP satellite. Its important to note that $\langle \sin(2\rho) \rangle$ is almost $0$ in all the pixels.}
\label{fig:cos2rho}
\end{figure} 
\begin{figure}
\centering
\includegraphics[width=0.5\textwidth,trim = 0 260 30 270, clip]{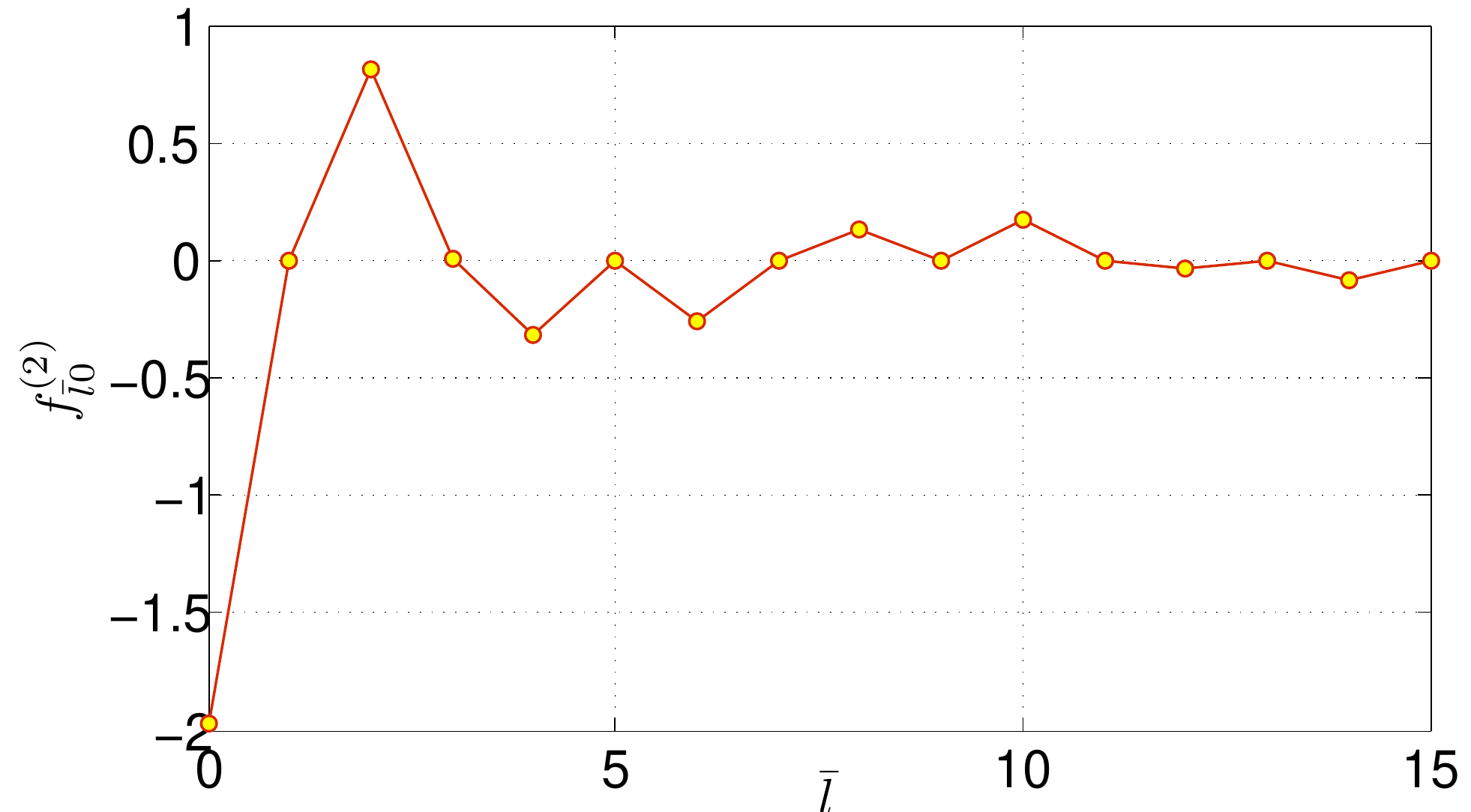} 
\caption{
Spherical harmonics of a $\cos{2\rho}$ map of WMAP scan are significantly dominated by $m=0$ modes. This is expected as WMAP scan is azimuthally symmetric in ecliptic coordinates. Also, the amplitude of SH coefficients of $\cos{2\rho}$ map decreases with increasing multipoles $l$ and having negligible contribution from odd multipoles $l$.
As can be seen in above figure, $l=0$ and $l=2$ modes are most dominant and are used for analytical estimation  effect of scan on beam-BipoSH. Nevertheless, the effect of scanning strategy can be evaluated by taking all significant SH coefficients into account for any experiment.}
\label{fig:scan-SH}
\end{figure}

\noindent The above equation gets simplified under the following assumptions :
\begin{itemize}
\item{}
The only dominant modes of $b_{l_2 m'}(\hat z)$ in Eq.(\ref{eq:genbeambiposh}) for the WMAP beam are the $|m'|=0,\ 2$ modes (see Sec.~\ref{BipoSHWMAPrawbeam}). Therefore, only ${\mathcal B}_{l_{1}l}^{LM^{(m')}}$ with $m'=0,\pm2$ will contribute to the above equation. Since the imaginary part of  $\chi_2(\hat n')$ is zero (see Fig.~\ref{fig:cos2rho}), it implies $\chi_2(\hat n') = \chi_{-2}(\hat n')$. 
\item{}
 The WMAP scan pattern is azimuthally symmetric about ecliptic pole axis. Therefore, in the ecliptic coordinate we can assume $f_{\bar{l}\bar{m}} = 0$ for all $\bar{m} \ne 0$. 
\end{itemize}
Under these assumptions Eq.(\ref{eq:ajk}) simplifies to

\begin{equation}
\tilde{a}_{JK}=\sum_{l_{1}\bar{l}}\frac{\Pi_{\bar{l}}\Pi_{l_{1}}}{\sqrt{4\pi}\Pi_{J}}C_{\bar{l}0l_{1}0}^{J0}C_{\bar{l}0l_{1}K}^{JK}
\left(\tilde{\mathfrak{A}}_{l_{1}K}^{(c)}f_{\bar{l}0}^{(0)}+\tilde{\mathfrak{A}}_{l_{1}K}^{(nc)}f_{\bar{l}0}^{(2)}\right)
\label{eq:ajkscanned}
\end{equation}
\noindent where, 
\begin{eqnarray}
\tilde{\mathfrak{A}}_{l_{1}m_{1}}^{(c)}&=&\sum_{lmLM}\left(-1\right)^{m}a_{lm}{\mathcal B}_{l_{1}l}^{LM^{(0)}}C_{l_{1}m_{1}l-m}^{LM} = \sum_{lm}\frac{\left(-1\right)^{m}}{\sqrt{2l+1}}a_{lm}{\mathcal B}_{ll}^{00}\,, \\
\tilde{\mathfrak{A}}_{l_{1}m_{1}}^{(nc)}&=&\sum_{lmLM}\sum_{m'=2,-2}\left(-1\right)^{m}a_{lm}{\mathcal B}_{l_{1}l}^{LM^{(m')}}C_{l_{1}m_{1}l-m}^{LM}\,.
\end{eqnarray}

\noindent $\tilde{\mathfrak{A}}_{l_{1}m_{1}}^{(c)}$, $\tilde{\mathfrak{A}}_{l_{1}m_{1}}^{(nc)}$
are the coefficients of the spherical harmonics of a sky convolved with the circular and
the non-circular part of the beam respectively, using PT scan. 
Beam-BipoSH coefficients  as obtained from the circular part of the beam ($m'=0$) i.e. ${\mathcal B}_{l_1 l}^{LM^{(0)}}$ are non vanishing only for $L=0$. Similarly the beam-BipoSH coefficients obtained from the non-circular part of the beam ($m' \ne 0$) are non zero only for $L\ne0$. Since the deviation from circularity is mild, ${\mathcal B}_{l_1 l}^{LM^{(0)}}$ will be dominant as compared to ${\mathcal B}_{l_1 l}^{LM^{(m'\ne 0)}}$.

Using Eq.(\ref{eq:ajkscanned}), the BipoSH coefficients from the scanned sky can be obtained as 

\begin{eqnarray}
\tilde{A}_{l_{1}l_{2}}^{L'M'} & = & \sum_{m_{1}m_{2}}\left\langle \tilde{a}_{l_{1}m_{1}}\tilde{a}_{l_{2}m_{2}}\right\rangle C_{l_{1}m_{1}l_{2}m_{2}}^{L'M'} \nonumber \\
 & = & \sum_{m_{1}m_{2}}\sum_{l\bar{l}}\sum_{l'\bar{l}'
}\frac{\Pi_{\bar{l}}\Pi_{l}\Pi_{\bar{l}'}\Pi_{l'}}{4\pi\Pi_{l_{1}}\Pi_{l_{2}}}C_{\bar{l}0l0}^{l_{1}0}C_{\bar{l}0lm_{1}}^{l_{1}m_{1}}C_{\bar{l}'0l'0}^{l_{2}0}C_{\bar{l}'0l'm_{2}}^{l_{2}m_{2}}\times
  \left[\left\langle \tilde{\mathfrak{A}}_{lm_{1}}^{(c)}\tilde{\mathfrak{A}}_{l'm_{2}}^{(c)}\right\rangle f_{\bar{l}0}^{(0)}f_{\bar{l}'0}^{(0)}\right. \nonumber \\
&&\left.+\left\langle \tilde{\mathfrak{A}}_{lm_{1}}^{(nc)}\tilde{\mathfrak{A}}_{l'm_{2}}^{(c)}\right\rangle f_{\bar{l}0}^{(2)}f_{\bar{l}'0}^{(0)}
+\left\langle \tilde{\mathfrak{A}}_{lm_{1}}^{(c)}\tilde{\mathfrak{A}}_{l'm_{2}}^{(nc)} \right\rangle f_{\bar{l}0}^{(0)}f_{\bar{l}'0}^{(2)}+\left\langle \tilde{\mathfrak{A}}_{lm_{1}}^{(nc)}\tilde{\mathfrak{A}}_{l'm_{2}}^{(nc)}\right\rangle f_{\bar{l}0}^{(2)}f_{\bar{l}'0}^{(2)}\right]\,.
\label{eq:Allp}
\end{eqnarray}

\begin{figure*}
\centering
\includegraphics[width=0.48\textwidth,trim = 0 260 20 280, clip]{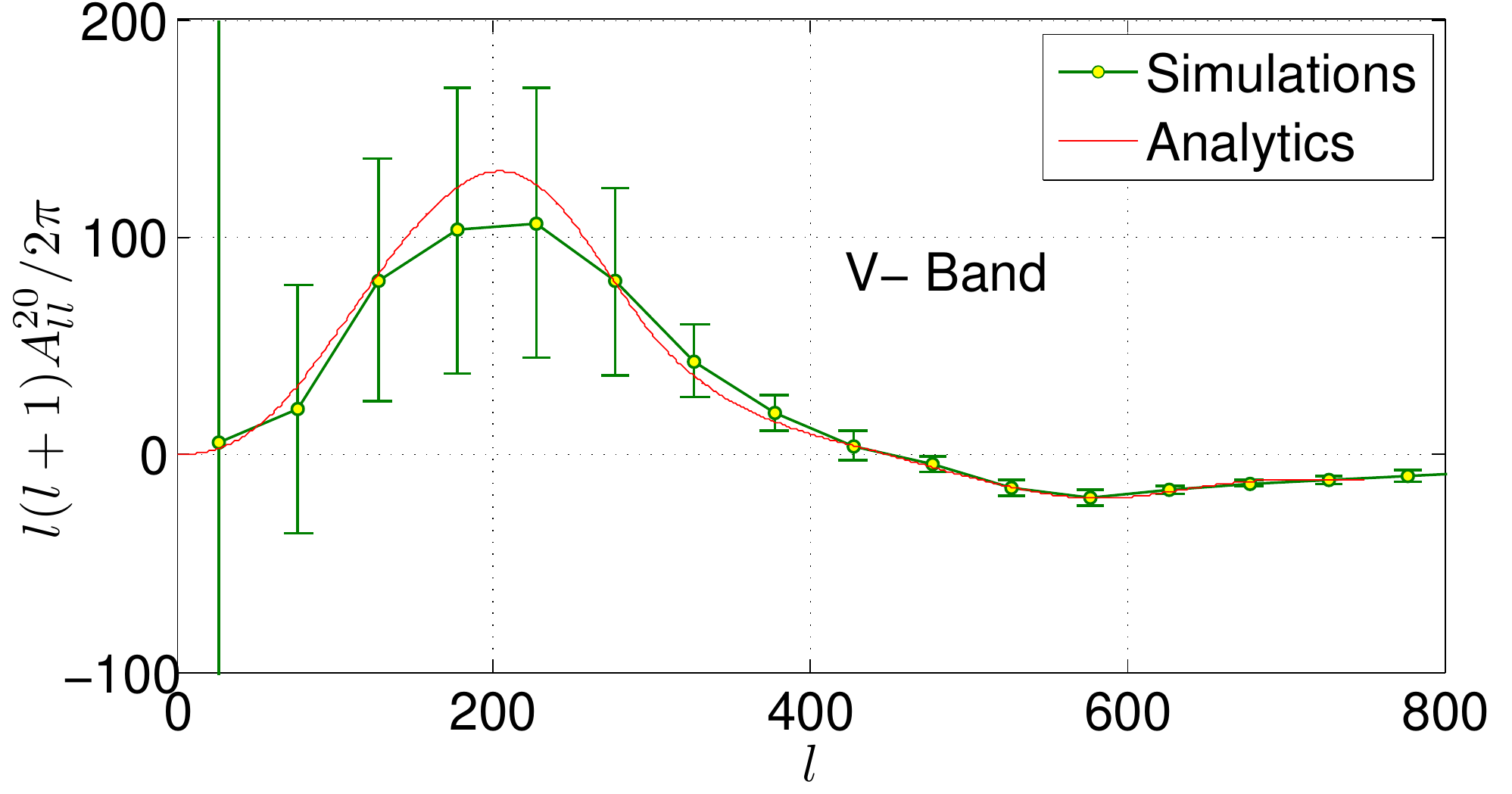}
\includegraphics[width=0.48\textwidth,trim = 0 260 20 280, clip]{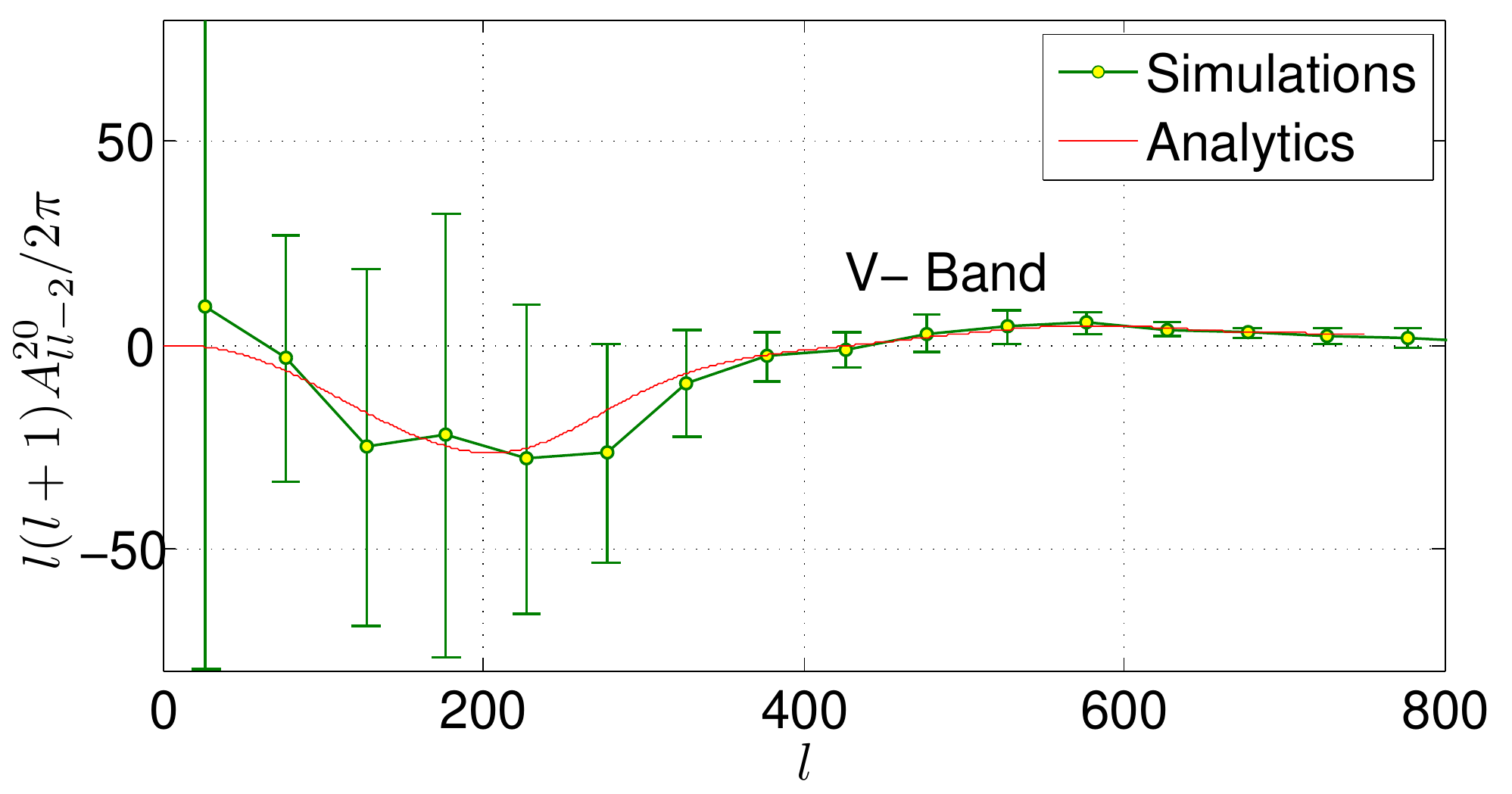}

\includegraphics[width=0.48\textwidth,trim = 0 260 20 280, clip]{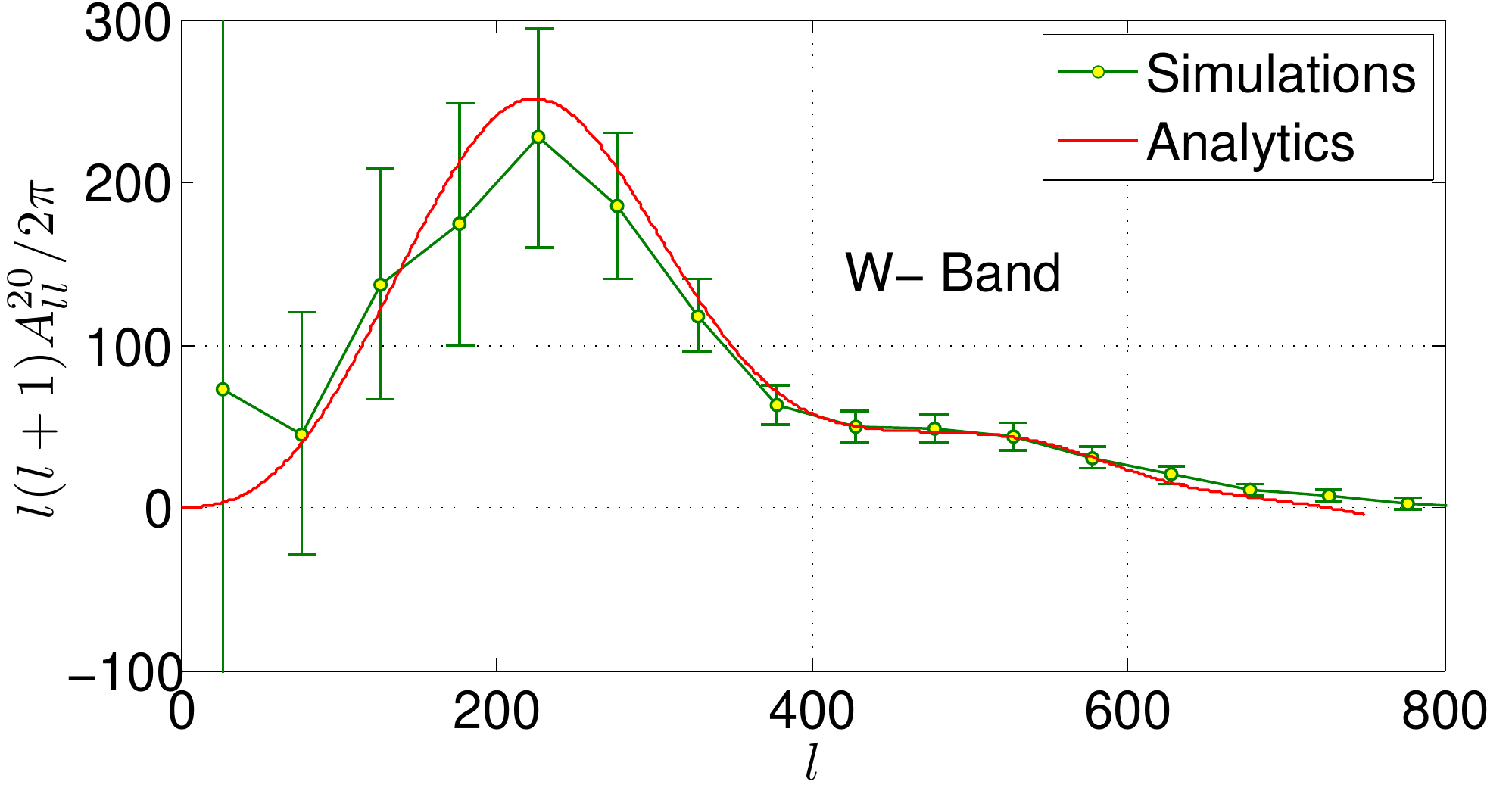}
\includegraphics[width=0.48\textwidth,trim = 0 260 20 280, clip]{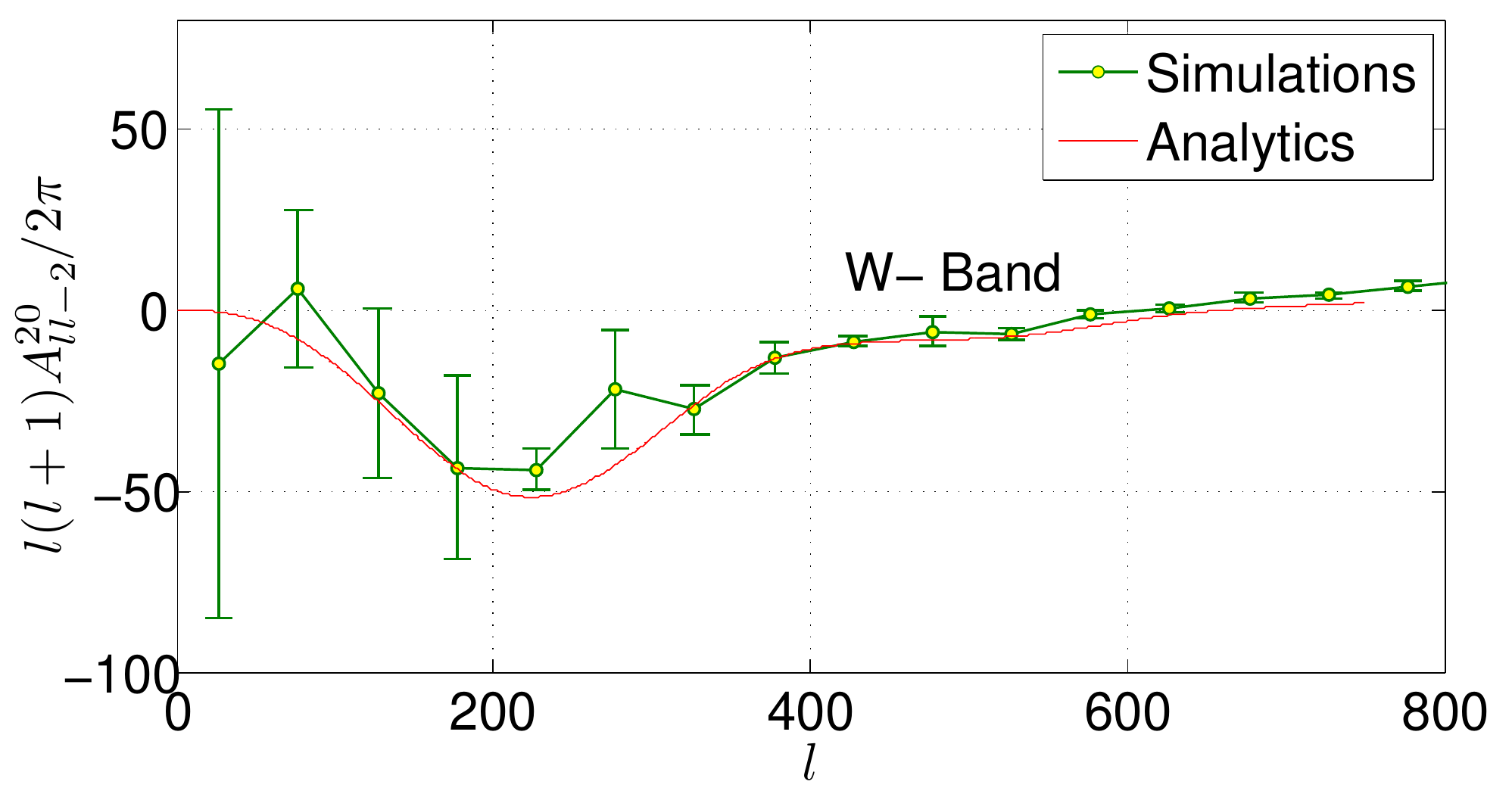}
\caption{BipoSH spectra, $A^{20}_{ll}$ ({\em Left}) and $A^{20}_{l-2
l}$ ({\em Right}) obtained for the raw beam maps A side of V2
channel ({\em Top}) and A side of W1 channel ({\em Bottom}) of the
WMAP experiment for the WMAP scan strategy. 
 Analytically evaluated BipoSH spectra (red)
overlaid on the average BipoSH spectra (green) with error bars obtained
from 30 simulations of statistically isotropic CMB sky convolved W1
and V2 channel of WMAP. Our approximate semi-analytical results matches 
very well with the numerical simulations.}
\label{fig:W1-V2fullblmscan}
\end{figure*}

\noindent The first term in the square bracket $\left\langle \tilde{\mathfrak{A}}_{lm_{1}}^{(c)}\tilde{\mathfrak{A}}_{l'm_{2}}^{(c)}\right\rangle$, being the covariance matrix of the spherical harmonic coefficients of the sky scanned by the circular part of the beam, can not contribute to $\tilde{A}_{l_{1}l_{2}}^{L'M'}$ for $L'\ne0$.
Since ${\mathcal B}_{l_1 l}^{LM^{(m'\ne 0)}}$ is sub-dominant, the last term $\left\langle \tilde{\mathfrak{A}}_{lm_{1}}^{(nc)}\tilde{\mathfrak{A}}_{l'm_{2}}^{(nc)}\right\rangle$ is 
negligible as compared to the rest of the terms.

We can expand the covariance matrix $\left\langle \tilde{\mathfrak{A}}_{lm_{1}}^{c}\tilde{\mathfrak{A}}_{l'm_{2}}^{nc}\right\rangle $,
in terms of the BipoSH spectra as

\begin{equation}
\left\langle \tilde{\mathfrak{A}}_{lm_{1}}^{c}\tilde{\mathfrak{A}}_{l'm_{2}}^{nc}\right\rangle =
\sum_{
LM}\tilde{A}_{ll'}^{LM^{(PT)}}C_{lm_{1}l'm_{2}}^{LM}\, \qquad\qquad\forall \,L\ne 0
\end{equation}

\noindent where $\tilde{A}_{ll'}^{LM^{(PT)}}$ are the BipoSH coefficients calculated
from a map scanned with PT scan.

Since $\chi_0(\hat n') =1$,  $f_{l0}^{(0)}=\sqrt{4\pi}\delta_{l0}$.
Also as discussed in the Sec.~\ref{bipforbeam}, $\tilde{A}_{ll'}^{LM^{(PT)}} =0$ for all $M\ne 0$ 
 in a coordinate system where PT scan is valid. 

Using the above details the Eq.(\ref{eq:Allp}) reduces to

\begin{equation}
\tilde{A}_{l_{1}l_{2}}^{L'0}=\sum_{l\bar{l}}\sum_{L\ne0}\tilde{A}_{ll_{2}}^{L0^{(PT)}}\left[\sum_{m}\frac{\Pi_{\bar{l}}\Pi_{l}}{\sqrt{4\pi}\Pi_{l_{1}}}C_{\bar{l}0l0}^{l_{1}0}C_{\bar{l}0lm_{1}}^{l_{1}m_{1}}C_{l_{1}m_{1}l_{2}-m_{1}}^{L'0}C_{lm_{1}l_{1}-m_{1}}^{L0}\right]f_{\bar{l}0}^{(2)}\,.
\end{equation}

\noindent Considering that the only BipoSH coefficients present in the parallel
transport scan with WMAP beam are $\tilde{A}_{ll}^{20^{(PT)}}$ and $\tilde{A}_{l-2l}^{20^{(PT)}}$
(as seen in the last section) we can obtain $\tilde{A}_{l_{1}l_{2}}^{L'0}$
for the proper WMAP scan as
\begin{eqnarray}
\tilde{A}_{ll}^{20}&=&(g^{0}_{l l}+g^{2}_{l l})\,\tilde{A}_{ll}^{20^{(PT)}}+\left(g^{2}_{l-2 l}\,\tilde{A}_{l-2l}^{20^{(PT)}}+g^{2}_{l+2 l}\,\tilde{A}_{l+2l}^{20^{(PT)}}\right) \label{eq:scAll}\\
\tilde{A}_{l-2l}^{20}&=&  g^{2}_{l-2 l}\,\tilde{A}_{ll}^{20^{(PT)}}+\left(g^{0}_{l-2 l} + g^{2}_{l-2 l-2} \right)\tilde{A}_{l-2 l}^{20^{(PT)}}\label{eq:scAll2}
\end{eqnarray}

 
\noindent where 
\begin{equation}
g^{\bar{l}}_{l_1 l} = \left[\sum_{m}\frac{\Pi_{\bar{l}}\Pi_{l}}{\sqrt{4\pi}\Pi_{l_{1}}}C_{\bar{l}0l0}^{l_{1}0}C_{\bar{l}0lm_{1}}^{l_{1}m_{1}}C_{l_{1}m_{1}l_{2}-m_{1}}^{20}C_{lm_{1}l_{1}-m_{1}}^{20}\right]f_{\bar{l}0}^{(2)}
\end{equation}

In Fig. \ref{fig:W1-V2fullblmscan} we have shown the plots that we obtain using WMAP beam and scan. We can see that the analytical results are matching almost exactly with the numerical simulations. 
For our calculations we first use an analytical approximation of the WMAP scan \cite{SD-TS2012a,Das2014} to calculate the scan angle $\rho_{\hat n j}$ at each scan point and obtain $\chi_2 (\hat n')$ map. The real and the imaginary part of the map are shown in Fig. \ref{fig:cos2rho}. 
The figure shows that the imaginary part of the $\chi_2 (\hat n')$ map is almost zero. The real part of the map, $\langle \cos(2\rho)\rangle$ is azimuthally symmetric. From this map we obtain the coefficients of the scan spherical harmonics, $f^{(2)}_{\bar l \bar m}$.  First $15$ modes of 
$f^{(2)}_{\bar l 0}$ are plotted in in Fig. \ref{fig:scan-SH}. With all these informations we obtain the BipoSH coefficients $A^{20}_{ll}$ and $A^{20}_{l-2l}$ using Eq.(\ref{eq:scAll}) and Eq.(\ref{eq:scAll2}). 

%
%
%
%
%

\section{Discussions \& Conclusion}\label{conclusions}

Current CMB experiments measures the temperature of the sky at finer angular resolution  and high sensitivity. 
Therefore, systematic effects have to be properly taken into account in the process of data analysis to 
consistently make cosmological inferences. The observed CMB sky is a convolution of the cosmological signal with
the instrumental beam response function of the experiment. The deconvolution of
the beam effect from the signal is relatively straightforward for an
ideal circularly symmetric beam. However, for a NC beam and complex scan, 
the deconvolution is practically impossible. Non-Circular (NC) deviations of the
beam, however mild, are practically inevitable in all experiments,
and affect the results obtained at the limits of the sensitivity and resolution of the recent experiments. 
CMB maps obtained with NC-beams and complex scan disrupt the
rotational invariance of the two point correlation function leading to clearly measurable signatures of 
SI violation.

We look for these SI violation signals in CMB measurement in BipoSH spectra. We introduce the novel and useful concept of expanding
the NC-beam response function in the BipoSH basis and refer them as beam-BipoSH coefficients. 
We investigate the impact of NC beam along with scan strategy on the beam-BipoSH and provide explicit analytical expressions for evaluating these
coefficients. Our approach is based on the harmonic expansion of the beam about the pointing direction and counting in fact that the power in $m$ modes decreases with increasing $m$ and odd $m$ modes are negligible in any realistic beam. We only take take first to two dominant modes $m=0$ and $m=2$ to evaluate our analytical expressions. 
We then obtain analytic expressions for observed CMB-BipoSH coefficients, which  incorporates non-circularity of the beam and scanning strategy, in terms of beam-BipoSH coefficients. 
To ease the complexity of the problem, we first obtain CMB-BipoSH coefficients generated by NC-beams along with a simplistic, idealized `parallel-transport' (PT) scan 
where the beam visits each pixel at a constant orientation ($\rho$). 
Then we extend our analytical formalism for a generalized scan, where the each sky pixel is observed multiple times and with a different orientation of the beam. 
The amplitude of the observed CMB-BipoSH coefficients for a PT scan is much higher as compared to a generalized scan strategy for an NC beam. This is expected as due to the multiple hits of the same pixel with different scan orientation tend to zero out the non-circular modes of the beam, thereby reducing the signature of the SI violation. 
Numerical simulations validate all our analytical expressions. We have taken WMAP to be an illustrative example for all our analysis. In particular our analytical estimates for a generalized scan fit well with the exact numerical simulation.

Exact numerical analysis for any experiment with the certain NC beam and scan strategy is immensely time consuming and takes 
tens of thousands of hours of CPU time on high-end clusters. 
On the contrary, our approximate semi-analytical method has an advantage of producing the results almost in no time yet recovering all the important 
features imprinted on the BipoSH coefficients due to the NC beam and the scan strategy. 
Our analytical expressions can be readily applied to any experiments to get estimates of CMB-BipoSH coefficients, 
 provided the beam has certain symmetries as discussed in the paper. 
This provides a new, powerful and efficient  machinery to address a rather complicated systematic effect of non-circular beam and scan strategy and to predict the level of SI violation for any given experiment. 
The analysis can be easily extended to study the effects of the beam on the CMB polarization maps. We defer this for future work.  

\section*{Acknowledgments}

We simulate CMB maps with the HEALPix~\cite{hpix} package additional
modules added for real-space convolution with NC-beams. Computations
were carried out at the HPC facilities at IUCAA.  SD
acknowledge the Council of Scientific and Industrial Research (CSIR),
India for financial support through Senior Research fellowships.

\appendix
\section{CMB ${\rm BipoSH}$ due to non-circular beams}\label{app:cmb-biposh}

Measured CMB temperature is a convolution of the instrumental beam response 
function and the underlying CMB temperature. Even if the
underlying cosmological temperature fluctuations are statistically
isotropic, non-circularity of the beam can give rise to detections in
BipoSH coefficients. 
The measured temperature fluctuation 
$\widetilde{\Delta T}(\hat n_{1})$ is given by 
\begin{eqnarray}\label{eq:convolution}
\widetilde{\Delta T}(\hat n_{1})=\int d{\Omega_{n_{2}}}B(\hat
n_{1},\hat n_{2})\Delta T(\hat n_{2}).
\end{eqnarray}
where $\Delta T(\hat n_{2})$ is the background sky temperature and  $B(\hat n_{1},\hat n_{2})$ is the beam
response function that encodes the
sensitivity of the instrument around the pointing direction, $\hat
n_1$. 

The CMB temperature field can be decomposed in the SH basis, as
\begin{eqnarray}
\Delta T(\hat n_2) =\sum_{lm}a_{lm}Y_{lm}(\hat n_2)\,.
\end{eqnarray}
Beam response function can be expanded in the BipoSH basis,
\begin{eqnarray}
B(\hat n_{1},\hat n_{2})=\sum_{l_1 l_2 L M}B^{LM}_{l_1 l_2}\sum_{m_1 m_2}C^{LM}_{l_1 m_1 l_2 m_2}\times Y_{l_1 m_1}(\hat n_{1})Y_{l_1 m_1}(\hat n_{2}).
\end{eqnarray}
Using orthogonality of spherical harmonics,
\begin{eqnarray}
\int d\Omega_{\hat n_2} Y_{lm}(\hat n_2)Y_{l'm'}(\hat n_2)=(-1)^{m'}\delta_{ll'}\delta_{mm'}\,,
\end{eqnarray}
we obtain
\begin{equation}
\widetilde{\Delta T}(\hat n_{1})=\sum_{l_1 m_1}\sum_{l m L
M}(-1)^{m}a_{lm}B^{LM}_{l_1 l}C^{LM}_{l_1 m_1 l -m} Y_{l_1 m_1}(\hat
n_{1}). 
\label{Tn1}
\end{equation}
This gives 
\begin{eqnarray}
\tilde a_{l_1 m_1}=\sum_{l m L M}(-1)^{m}a_{lm}B^{LM}_{l_1 l}C^{LM}_{l_1 m_1 l -m}\,,
\end{eqnarray}
where $\tilde a_{l_1 m_1}$ are the coefficients of the spherical harmonics expansion of $\widetilde{\Delta T}(\hat n_{1})$. The covariance matrix of these spherical harmonic coefficients can be calculated as 
\begin{eqnarray}\label{eq:1}
\langle \tilde a_{l_1 m_1}\tilde a_{l_2 m_2}\rangle \ = \ \sum_{lmLM}\sum_{l'm'L'M'} (-1)^{m+m'}\langle a_{l m}a_{l'
m'}\rangle 
 \times \ B^{LM}_{l_1 l}B^{LM}_{l_2 l'}C^{LM}_{l_1 m_1 l
-m}C^{L'M'}_{l_2 m_2 l' -m'}\,.
\end{eqnarray}
Assuming the CMB signal to be statistically isotropic, i.e.
\begin{eqnarray}
\langle a_{l m}a_{l' m'}\rangle=(-1)^{m}C_{l}\delta_{l l'}\delta_{m -m'}\,,
\end{eqnarray}
and substituting it in Eq.(\ref{eq:1}), we obtain the SH-space covariance
of the observed map as
\begin{eqnarray}
\langle \tilde a_{l_1 m_1}\tilde a_{l_2 m_2}\rangle \ = 
\sum_{lmLM}\sum_{L'M'} (-1)^{m}C_{l} B^{LM}_{l_1 l}B^{L'M'}_{l_2 l}C^{LM}_{l_1 m_1 l -m}C^{L'M'}_{l_2 m_2 l m}
\end{eqnarray}
Using Eq.(\ref{eq:gen-BipoSH}), we can calculate CMB-BipoSH coefficients as 
\begin{eqnarray}
\tilde A^{L_1 M_{1}}_{l_1 l_2} &=& \sum_{l L L' M M'} C_{l} B^{LM}_{l_1 l}B^{L'M'}_{l_2 l}\ \times
\sum_{m m_1 m_2}(-1)^{m} C^{LM}_{l_1 m_1 l -m} C^{L' M'}_{l_2 m_2 l m} C^{L_1 M_1}_{l_1 m_1 l_2 m_2} 
\end{eqnarray}
The sum over product of three Clebsch-Gordan coefficients can be
written compactly in terms of a $6$-j symbol, as
\begin{equation}
\sum_{\alpha\beta\delta}(-1)^{a-\alpha}C^{c\gamma}_{a\alpha
b\beta}C^{e\epsilon}_{d\delta b\beta}C^{f\varphi}_{d\delta a
-\alpha}=K_1\prod_{cf}C^{e\epsilon}_{c\gamma f\varphi}
{\begin{Bmatrix} a & b & c \\ e & f & d
\end{Bmatrix}}\,,
\end{equation}
where $K_1 =(-1)^{b+c+d+f}$ and  $\prod_{cf}=\sqrt{(2c+1)(2f+1)}$.  Hence, we obtain the
expression of Eq.(\ref{biposhbeam}) for CMB-BipoSH coefficient from
NC-beam,
 \begin{eqnarray}
\tilde A^{L_1 M_1}_{l_1 l_2} = \sum_{lLM L'M'} C_{l} \, B^{L M}_{l_1 l}\, B^{L' M'}_{l_2 l} \, (-1)^{l_1+L'-L_{1}}
  \sqrt{(2L+1)(2L'+1)}C^{L_1 M_1}_{L M L' M'}
{\begin{Bmatrix}
l & l_1 & L \\
L_1 & L' & l_2  
\end{Bmatrix}}
\,.
\end{eqnarray}
If we assume a PT-scan, then in that coordinate $M=0,M'=0, M_{1}=0$ (see Eq.~\ref{beamBiposhm2}). Thus the above expression reduces to,
 \begin{eqnarray}
\tilde A^{L_1 0}_{l_1 l_2} = \sum_{lL L'} C_{l} \, B^{L 0}_{l_1 l}\, B^{L' 0}_{l_2 l} \, (-1)^{l_1+L'-L_{1}} \times
 \sqrt{(2L+1)(2L'+1)}C^{L_1 0}_{L 0 L' 0}
{\begin{Bmatrix}
l & l_1 & L \\
L_1 & L' & l_2  
\end{Bmatrix}}
\,.
\end{eqnarray}
The Clebsch-Gordan coefficient $C^{L_1 0}_{L 0 L' 0}$ is zero when the
sum $L+L'+L_{1}$ is odd valued. Hence, it enforces the condition that the
summation in the above expression is limited to $L+L'+L_{1}$ being
even-valued. If the beam function has an even fold azimuthal symmetry ($m$ even in $b_{lm}(\hat n)$) and reflection symmetry ($l+m$ even in $b_{lm}(\hat n)$)  beam-BipoSH coefficients are restricted to even parity and follows 
$l_{1}+l_{2}=\textrm{even}$, then $L$ and $L'$ are restricted to even multipole values. Thereafter, due to the presence of $C^{L_1 0}_{L 0 L' 0}$, $L_1$ takes up even 
multipole values. 

\section{Beam BipoSH}\label{app:beam-biposh}

Beam-BipoSH are expansion coefficients of the beam response function
in BipoSH basis (see Sec~\ref{ncbeambposh}).  The most general
beam-BipoSH in any coordinate system is given by, 
\begin{eqnarray}
B^{LM}_{l_1 l_2} \ =\ -\sum_{m_1 m_2}C^{LM}_{l_1 m_1 l_2 m_2}\sum_{m'}b_{l_2
m'}(\hat z) \ \times
\int^{\pi}_{0} d(\cos\theta) \int^{2\pi}_0 d\phi D^{l_2}_{m_2
m'}(\phi,\theta,\rho(\theta,\phi))Y^{*}_{l_1
m_1}(\theta,\phi). \label{B1}
\end{eqnarray}
 Wigner-\textit{D} functions can be expressed in terms of Wigner-\textit{d} through following relation,
\begin{eqnarray}
D^{l}_{m m'}(\phi,\theta,\rho)={\rm e}^{-i m \phi}\, d^{l}_{m
m'}(\theta)\,{\rm e}^{-i m' \rho}\,.\label{B2}
\end{eqnarray}
and reduces to spherical harmonics for $m'=0$,
\begin{eqnarray}
D^{l}_{m0}(\phi,\theta,\rho)=\sqrt{4\pi/(2l+1)}Y^{*}_{lm}(\theta,\phi)\,.\label{B3}
\end{eqnarray}

In the parallel-transport (PT) scan, the beam
orientation, with respect to the local Cartesian coordinate aligned
with the spherical $(\hat{\theta}, \hat{\phi})$ coordinates, does not vary on sky
(i.e., $\rho(\theta,\phi)\equiv\rho$). Substituting Eq.(\ref{B2}) and  Eq.(\ref{B3}) into Eq.(\ref{B1}), and after some algebraic manipulation 
we get the beam-BipoSH coefficients for PT scan as,

\begin{eqnarray}
B^{LM}_{l_1 l_2}&=&
2\pi\delta_{M0}\sum_{m'}b_{l_2 m'}(\hat z) e^{-i m' \rho} \times \sum_{m_2}(-1)^{m_2}C^{L0}_{l_1 -m_2 l_2 m_2} I^{l_1 l_2}_{m_2,m'},
\end{eqnarray}
The beam-BipoSH coefficients are non-zero only for $M=0$, which originally comes from the relation,

\begin{eqnarray}
\int^{2\pi}_0 d\phi\exp^{-i (m_1+m_2) \phi} \ = \ 2\pi\delta_{m_1,-m_2}.
\end{eqnarray}
The notation $ I^{l_1 l_2}_{m_2,m'}$ is defined as 
\begin{eqnarray}
 I^{l_1 l_2}_{m_2,m'}=(-1)^{m_2 +1} \sqrt{\frac{(2l_{1}+1)}{4\pi}}\times \int^{\pi}_{\theta=0}d^{l_2}_{m_{2} m'}(\theta)d^{l_1}_{m_{2} 0}(\theta)d(\cos\theta).
\end{eqnarray}
Here we use the relation $d^{l}_{m m'}(\theta)=(-1)^{m-m'}d^{l}_{-m -m'}(\theta)$.

To simplify the analytic expressions, we retain only the leading order
NC beam spherical harmonic mode $m'=2$, assuming mild NC-beam with
discrete even-fold azimuthal symmetry where no odd $m'$ modes will
contribute. Hence, the summation over $m'$ has three terms, 
$m'=0,\pm2$.

The beam-BipoSH can be then be written as
\begin{eqnarray}\label{eq:thterm}
B^{LM}_{l_1 l_2} &\equiv & B^{LM^{(C)}}_{l_1 l_2}+B^{LM^{(NC)}}_{l_1 l_2}\,, \\
B^{LM^{(C)}}_{l_1 l_2} &=& 2\pi\,\delta_{L0}\,\delta_{M0} \ b_{l_2 0}(\hat z)\sum_{m_2}C^{L0}_{l_1 -m_2 l_2 m_2} I^{l_1
l_2}_{m_2,0}\,, \label{B8}\\
B^{LM^{(NC)}}_{l_1 l_2} &=& 2\pi\delta_{M0} \sum_{m_{2}\neq 0}C^{L0}_{l_1 -m_2 l_2 m_2} \times 
\Big( b_{l_2 -2}(\hat z)\exp^{i2\rho} I^{l_1 l_2}_{m_2,-2}+ b_{l_2 2}(\hat z)~\exp^{-i2\rho}~ I^{l_1 l_2}_{m_2,2} \Big). 
\end{eqnarray}
First term in Eq.(\ref{eq:thterm}) is the trivial beam-BipoSH,
$B^{00^{(C)}}_{l l}$, corresponding to the circular symmetric component of
the beam response function.  NC part of the beam function
$m'=\pm2$, gives rise to beam BipoSH having $L\neq0$. 

\subsection{Evaluating the circular part of beam-BipoSH coefficients}
First, we
evaluate the beam-BipoSH due to circular part of beam function.
Orthogonality of Wigner-\textit{d} functions,
\begin{eqnarray}
-\int^{\pi}_{0} d(\cos\theta) \,d^{l}_{m m'}(\theta)d^{l'}_{m m'}(\theta)=\frac{2}{2l+1}\delta_{l l'}
\end{eqnarray}
implies 
\begin{eqnarray}
I^{l_1 l_2}_{m_2,0}=(-1)^{m_2}\Big(\frac{2}{2l_2+1}\Big)\sqrt{\frac{(2l_{1}+1)}{4\pi}}\delta_{l_1 l_2}. \label{B11}
\end{eqnarray}

\noindent Substituting  Eq.(\ref{B11}) into Eq.(\ref{B8}) and using the property of the Clebsch-Gordan coefficients,
$\sum_{m}(-1)^{l-m}C^{L0}_{l m l -m}=\sqrt{(2l+1)}\delta_{L0}$\,,
we obtain
\begin{eqnarray}
B^{LM^{(C)}}_{l_1 l_2}=\sqrt{4\pi}(-1)^{l_2}b_{l_{2} 0}(\hat z)\delta_{l_1 l_2}\delta_{L0}\delta_{M0}.
\end{eqnarray}
Since,
\begin{eqnarray}
b_{l0}(\hat z)= \sqrt{\frac{(2l+1)}{4\pi}}B_{l},
\end{eqnarray}
where $B_{l}$ is the usual beam transfer function of the
circular-symmetrized beam profile,

\begin{eqnarray}
B^{LM(C)}_{l_1 l_2}=(-1)^{l_2}\sqrt{2l_{2}+1}B_{l_2}\delta_{l_1 l_2}\delta_{L0}\delta_{M0}. 
\end{eqnarray}

\subsection{Evaluating the non-circular part of the beam-BipoSH coefficients}
The NC part of the beam-BipoSH coefficient ($B^{LM^{(NC)}}$) is given by 
\begin{eqnarray}
B^{LM^{(NC)}}_{l_1 l_2} &=&-2\pi\sqrt{\frac{(2l_{1}+1)}{4\pi}}\delta_{M0}\sum_{m_2}(-1)^{m_2}C^{L0}_{l_1 -m_2 l_2 m_2}
\times \Big(b_{l_2 -2}(\hat z)\exp^{i2\rho}
\int^{\pi}_{0}d^{l_2}_{m_{2} -2}(\theta)d^{l_1}_{m_{2} 0}(\theta)
 d(\cos\theta) \nonumber \\
&&+ \ b_{l_2 2}(\hat z)\exp^{-i2\rho}\int^{\pi}_{0}d^{l_2}_{m_{2} 2}(\theta)d^{l_1}_{m_{2} 0}(\theta)d(\cos\theta)
\Big)\,.
\label{eq:NC-BipoSH}
\end{eqnarray}
In the above expression, the summation is over $m_2$. It is convenient
to separate the calculation of the $m_2=0$ and rest of the
$m_{2}\neq0$ terms.

\vspace{1em}
\noindent {\bf Calculating the ${\rm m_2 = 0}$ term :}
\newline \newline
Consider the integral for $m_2=0$.  Using the relations, $d^{l}_{m m'}=(-1)^{m+m'}d^{l}_{m' m}$ and expansion of Wigner-\textit{d}'s in
terms of associated Legendre polynomials, $d^{l}_{m
0}(\theta)=(-1)^{m}\sqrt{(l-m)!/(l+m)!}P^{m}_{l}(\cos\theta)$, we can 
obtain
\begin{eqnarray}
\int^{\pi}_{\theta=0}d^{l_2}_{0 2}(\theta)d^{l_1}_{0
0}(\theta)d(\cos\theta) \ = 
\sqrt{\frac{(l_{2}-2)!}{(l_{2}+2)!}}
\int^{\pi}_{\theta=0}P^{2}_{l_2}(\cos\theta)P_{l_1}(\cos\theta)d(\cos\theta)\,. \nonumber
\end{eqnarray}
where $P_{l_1}(\cos\theta)$ is the Legendre polynomial. Using standard recurrence relations of Associated Legendre functions,
\begin{eqnarray}
P^{2}_{l}(\cos\theta)&=&\frac{2\cos\theta}{\sin\theta}P^{1}_{l}(\cos\theta)-l(l+1)P_{l}(\cos\theta)\,, \\
P^{1}_{l}(\cos\theta)&=&\sin\theta P^{\prime}_{l}(\cos\theta) ,
\end{eqnarray}
and orthogonality relations,
\begin{eqnarray}
-\int^{\pi}_{0}
P_{l_2}(\cos\theta)P_{l_1}(\cos\theta)d(\cos\theta) \ &=&\ \frac{2~\delta_{l_1
l_2}}{2l_{2}+1}\,, \\
-\int^{\pi}_{0}\cos\theta P^{{\prime}}_{l_2}(\cos\theta)P^{0}_{l_1}(\cos\theta)d(\cos\theta)  
&=& \ \left\{
\begin{array}{ll}
 0 & \mbox{if ($l_1 +l_2=\textrm{odd}$)} \\ 
0 & \mbox{if ($l_1 >l_2$)} \\ 
0 & \mbox{if ($l_1 <l_2$)} \\ 
\frac{2l_{2}}{2l_{2}+1} & \mbox{if ($l_1 = l_2$)}  \,\\
\end{array}
\right.
\end{eqnarray}
the integral for $m_2=0$, simplifies to
\begin{eqnarray}
I^{l_1 l_2}_{0,\pm2}\ \ = \ (-1)^{m_2}\sqrt{\frac{(2l_1 +1)}{4\pi}}\times 
\left\{
\begin{array}{ll}
0 & \mbox{if ($l_1 +l_2=\textrm{odd}$)} \\
0 & \mbox{if ($l_1 >l_2$)} \\
4\sqrt{\frac{(l_{2}-2)!}{(l_{2}+2)!}} & \mbox{if ($l_1 <l_2$)} \\
\sqrt{\frac{(l_{2}-2)!}{(l_{2}+2)!}}\big[\frac{4l_2}{(2l_2 +1)}-\frac{2l_2(l_2 +1)}{(2l_2 +1)}\big] & \mbox{if ($l_1 = l_2$)} \,.\\ 
\end{array}
\right.
\end{eqnarray}

\vspace{1em}
\noindent {\bf Calculating the $m_2 \ne 0$ term : }
\newline \newline
Next, we evaluate $m_2\neq0$ terms in the summation in
Eq.(\ref{eq:NC-BipoSH}).  $d^{l_2}_{m_{2} 2}(\theta)$ can be
recursively reduced to $d^{l_2}_{m_{2} 0}(\theta)$ using the following
recurrence relation,
\begin{eqnarray}\label{d1}
d^{l_2}_{m_{2} 2}(\theta) = \frac{\kappa}{\sin^{2}\theta}\left[\kappa_{0}d^{l_2}_{m_2 0}(\theta) + \kappa_{1}d^{l_{2}+1}_{m_2 0}(\theta) \ + \kappa_{-1}d^{l_{2}-1}_{m_2 0}(\theta)+\kappa_{2}d^{l_{2}+2}_{m_2 0}(\theta)+\kappa_{-2}d^{l_{2}-2}_{m_2 0}(\theta)\right]\,.
\end{eqnarray}
where,
\begin{eqnarray*} 
\kappa_0 \ &\equiv& \ \frac{m^2_{2}}{l^2_{2}(l_{2}+1)^2} \ -\
\frac{l_{2}^2-m^2_{2}}{l_{2}^2(4l^2_{2}-1)} 
 - \ \frac{(l_{2}+1)^2-m^2_{2}}{(l_{2}+1)^2(2l_{2}+1)(2l_{2}+3)}, \\ \kappa_1 \ &\equiv& \
2m_{2}\frac{\sqrt{(l_{2}+1)^2-m^2_{2}}}{l_{2}(l_{2}+1)(l_{2}+2)(2l_{2}+1)}, \\
\kappa_{-1} &\equiv&  -2m_{2}\frac{\sqrt{l^2_{2}-m^2_{2}}}{l_{2}(l^2_{2}-1)(2l_{2}+1)},\\ \kappa_2 \
&\equiv& \
\frac{\sqrt{[(l_{2}+1)^2-m^2_{2}][(l_{2}+2)^2-m^2_{2}]}}{(l_{2}+1)(l_{2}+2)(2l_{2}+1)(2l_{2}+3)} \\
\kappa_{-2} &\equiv &
\frac{\sqrt{(l_{2}^2-m^2_{2})[(l_{2}-1)^2-m^2_{2}]}}{l_{2}(l_{2}-1)(4l^2_{2}-1)} \,.
\end{eqnarray*}
Under reflection symmetry, the Wigner-\textit{d}'s transform as,
$d^{l}_{m m'}(\pi-\theta)=(-1)^{l+m'}d^{l}_{m -m'}(\theta)$. Using this we obtain
\begin{eqnarray}\label{d2}
d^{l_2}_{m_{2} -2}(\theta) =\frac{\kappa}{\sin^{2}\theta}\left[\kappa_{0}d^{l_2}_{m_2 0}(\theta) -\kappa_{1}d^{l_{2}+1}_{m_2 0}(\theta)  - \ \kappa_{-1}d^{l_{2}-1}_{m_2 0}(\theta) +\kappa_{2}d^{l_{2}+2}_{m_2 0}(\theta)+\kappa_{-2}d^{l_{2}-2}_{m_2 0}(\theta) \right].
\end{eqnarray}
Substituting Eq.(\ref{d1}) and Eq.(\ref{d2}) in Eq.(\ref{eq:NC-BipoSH}) and 
using the relation~\cite{SM-AS-TS},  
\begin{eqnarray}
\int^{2\pi}_{0} \frac{d^{l_2}_{m_2 0}(\theta) d^{l_2}_{m_2 0}(\theta)}{\sin^{2}(\theta)}d\theta
&=&\left\{
\begin{array}{ll}
\frac{1}{m_2}\sqrt{\frac{(l_{2}-|m_2|)!(l_{1}+|m_2|)!}{(l_{2}+|m_2|)!(l_{1}-|m_2|)!}}\quad\quad\quad l_{1}< l_{2}\nonumber\\ 
\frac{1}{m_2}\sqrt{\frac{(l_{2}+|m_2|)!(l_{1}-|m_2|)!}{(l_{2}-|m_2|)!(l_{1}+|m_1|)!}}\quad\quad\quad l_{1}> l_{2}\nonumber\\
 \frac{1}{m_2}\;\;\quad\quad\quad\quad\quad\quad\quad\quad\quad\quad\quad l_{1}= l_{2}\,
\end{array}
\right.
\end{eqnarray}
the expressions for  $I^{l_1 l_2}_{m_2,\pm2}$  for $m_2\neq0$ becomes 
\begin{eqnarray}
I^{l_1 l_2}_{m_2,\pm2}&=&  (-1)^{m_2}\sqrt{\frac{(2l_1 +1)}{4\pi}}\ \left\{
\begin{array}{ll}
\Big(\ \frac{\kappa \kappa_{0}}{|m_2|}+
\frac{\kappa \kappa_{2}}{|m_2|}\sqrt{\frac{(l_{2}+|m_2|)!(l_{2}+2-|m_2|)!}{(l_{2}-|m_2|)!(l_{2}+2+|m_2|)!}}\nonumber \\
+\frac{\kappa \kappa_{-2}}{|m_2|}\sqrt{\frac{(l_{2}-|m_2|)!(l_{2}-2+|m_2|)!}{(l_{2}+|m_2|)!(l_{1}-2-|m_2|)!}}\ \Big) & \mbox{if ($l_1 =l_2$)} \\ \\ \\
\Big(\ \frac{\kappa \kappa_{0}}{|m_2|}\sqrt{\frac{(l_{2}+|m_2|)!(l_{1}-|m_2|)!}{(l_{2}-|m_2|)!(l_{1}+|m_2|)!}}+
\frac{\kappa \kappa_{2}}{|m_2|}\sqrt{\frac{(l_{2}+2+|m_2|)!(l_{1}-|m_2|)!}{(l_{2}+2-|m_2|)!(l_{1}+|m_2|)!}}
+\nonumber\\\frac{\kappa \kappa_{-2}}{|m_2|}\sqrt{\frac{(l_2-2+|m_2|)!(l_1-|m_2|)!}{(l_2-2-|m_2|)!(l_1+|m_2|)!}}\pm
\frac{\kappa \kappa_{1}}{|m_2|}\sqrt{\frac{(l_{2}+1+|m_2|)!(l_{1}-|m_2|)!}{(l_{2}+1-|m_2|)!(l_{1}+|m_2|)!}}
\nonumber\\\pm\frac{\kappa \kappa_{-1}}{|m_2|}\sqrt{\frac{(l_{2}-1+|m_2|)!(l_1-|m_2|)!}{(l_{2}-1-|m_2|)!(l_1+|m_2|)!}}\ \Big) & \mbox{if ($l_1 >l_2$)} \\ \\ \\
\Big(\ \frac{\kappa \kappa_{0}}{|m_2|}\sqrt{\frac{(l_{1}+|m_2|)!(l_{2}-|m_2|)!}{(l_{1}-|m_2|)!(l_{2}+|m_2|)!}}+
\frac{\kappa \kappa_{2}}{|m_2|}\sqrt{\frac{(l_{2}+2-|m_2|)!(l_{1}+|m_2|)!}{(l_{2}+2+|m_2|)!(l_{1}-|m_2|)!}}\nonumber\\
+\frac{\kappa \kappa_{-2}}{|m_2|}\sqrt{\frac{(l_{2}-2-|m_2|)!(l_1+|m_2|)!}{(l_{2}-2+|m_2|)!(l_1-|m_2|)!}}\pm
\frac{\kappa \kappa_{1}}{|m_2|}\sqrt{\frac{(l_{2}+1-|m_2|)!(l_{1}+|m_2|)!}{(l_{2}+1+|m_2|)!(l_{1}-|m_2|)!}}
\nonumber\\ \pm\frac{\kappa \kappa_{-1}}{|m_2|}\sqrt{\frac{(l_{2}-1-|m_2|)!(l_1+|m_2|)!}{(l_{2}-1+|m_2|)!(l_1-|m_2|)!}}\ \Big) & \mbox{if ($l_1 <l_2$)}\,. \\ 
\end{array}
\right.
\end{eqnarray}

In general, NC-beams would generate both even-parity (+) and odd-parity (-)
beam-BipoSH coefficients
\begin{eqnarray}\label{eq:oddparity-BipoSH}
B^{LM^{(NC)}}_{l_1 l_2}=B^{LM^{(+)}}_{l_1 l_2}+B^{LM^{(-)}}_{l_1 l_2}\,.
\end{eqnarray}

\noindent The even-parity beam-BipoSH,
\begin{eqnarray}
B^{LM^{(+)}}_{l_1 l_2}&=&\left\{ 
\begin{array}{ll}
 \delta_{M0}\left[b_{l_2 2}(\hat z)\exp({-i2\rho})+b^{*}_{l_2 2}(\hat z)\exp({i2\rho})\right] \times \nonumber \\
\left(-C^{L0}_{l_1 0 l_2 0}\sqrt{\frac{4\pi l_2 (l_{2}-1)}{(2l_{2}+1)(l_{2}+2)(l_{2}+1)}}+ 
2\pi\sqrt{\frac{(2l_{1}+1)}{4\pi}}\sum_{|m_2|>0}C^{L0}_{l_1 -m_2 l_2 m_2} \right.\times\nonumber\\
\big[ \frac{\kappa \kappa_{0}}{|m_2|}+
\frac{\kappa \kappa_{2}}{|m_2|}\sqrt{\frac{(l_{2}+|m_2|)!(l_{2}+2-|m_2|)!}{(l_{2}-|m_2|)!(l_{2}+2+|m_2|)!}} 
\left.+\frac{\kappa \kappa_{-2}}{|m_2|}\sqrt{\frac{(l_{2}-|m_2|)!(l_{2}-2+|m_2|)!}{(l_{2}+|m_2|)!(l_{1}-2-|m_2|)!}}\big]\right) & \mbox{if ($l_1 =l_2$ and $l_2\geq2$)} \nonumber\\ \\ \\

 \delta_{M0}\left[b_{l_2 2}(\hat z)\exp({-i2\rho})+b^{*}_{l_2 2}(\hat z)\exp({i2\rho})\right]2\pi\sqrt{\frac{(2l_{1}+1)}{4\pi}}\times \nonumber\\
\left(\sum_{|m_2|>0}C^{L0}_{l_1 -m_2 l_1 m_2}\right.\times
\big[ \frac{\kappa \kappa_{0}}{|m_2|}\sqrt{\frac{(l_{2}+|m_2|)!(l_{1}-|m_2|)!}{(l_{2}-|m_2|)!(l_{1}+|m_2|)!}}+ \nonumber\\
\frac{\kappa \kappa_{2}}{|m_2|}\sqrt{\frac{(l_{2}+2+|m_2|)!(l_{1}-|m_2|)!}{(l_{2}+2-|m_2|)!(l_{1}+|m_2|)!}}
+\left.\frac{\kappa \kappa_{-2}}{|m_2|}\sqrt{\frac{(l_2-2+|m_2|)!(l_1-|m_2|)!}{(l_2-2-|m_2|)!(l_1+|m_2|)!}}\big]\right) & \mbox{if ($l_1 >l_2$ and $l_2\geq2$)} \nonumber\\ \\ \\

\delta_{M0}[b_{l_2 2}(\hat z)\exp({-i2\rho})+b^{*}_{l_2 2}(\hat z)\exp({i2\rho})]\left(8\pi C^{L0}_{l_1 0 l_2 0}\sqrt{\frac{(2l_{1}+1)(l_{2} -2)!}{4\pi(l_{2}+2)!}}+\right.\nonumber\\
2\pi\sqrt{\frac{(2l_{1}+1)}{4\pi}}\sum_{|m_2|>0}C^{L0}_{l_1 -m_2 l_1 m_2}\big[ \frac{\kappa \kappa_{0}}{|m_2|}\sqrt{\frac{(l_{1}+|m_2|)!(l_{2}-|m_2|)!}{(l_{1}-|m_2|)!(l_{2}+|m_2|)!}}+\nonumber\\
\frac{\kappa \kappa_{2}}{|m_2|}\sqrt{\frac{(l_{2}+2-|m_2|)!(l_{1}+|m_2|)!}{(l_{2}+2+|m_2|)!(l_{1}-|m_2|)!}}
+\left.\frac{\kappa \kappa_{-2}}{|m_2|}\sqrt{\frac{(l_{2}-2-|m_2|)!(l_1+|m_2|)!}{(l_{2}-2+|m_2|)!(l_1-|m_2|)!}}\big]\right) & \mbox{if ($l_1 <l_2$ and $l_2\geq2$)}\nonumber\,,
\end{array}
\right.
\end{eqnarray}
and odd parity-beam-BipoSH,
\begin{eqnarray}
B^{LM^{(-)}}_{l_1 l_2}=\left\{ 
\begin{array}{ll}
0 & \mbox{if ($l_1 = l_2$)} \nonumber\\ \\

 \delta_{M0}\left[b_{l_2 2}(\hat z)\exp({-i2\rho})-b^{*}_{l_2 2}(\hat z)\exp({i2\rho})\right]2\pi\sqrt{\frac{(2l_{1}+1)}{4\pi}}\times \nonumber\\
\left(\sum_{|m_2|>0}C^{L0}_{l_1 -m_2 l_1 m_2} \times\big[
\frac{\kappa \kappa_{1}}{|m_2|}\sqrt{\frac{(l_{2}+1+|m_2|)!(l_{1}-|m_2|)!}{(l_{2}+1-|m_2|)!(l_{1}+|m_2|)!}} \right.\nonumber\\
+\left.\frac{\kappa \kappa_{-1}}{|m_2|}\sqrt{\frac{(l_{2}-1+|m_2|)!(l_1-|m_2|)!}{(l_{2}-1-|m_2|)!(l_1+|m_2|)!}}\big]\right) & \mbox{if ($l_1 >l_2$ and $l_2\geq2$)} \nonumber\\ \\

 \delta_{M0}[b_{l_2 2}(\hat z)\exp({-i2\rho})-b^{*}_{l_2 2}(\hat z)\exp({i2\rho})]2\pi\sqrt{\frac{(2l_{1}+1)}{4\pi}} \times  \nonumber\\
\left(\sum_{|m_2|>0}C^{L0}_{l_1 -m_2 l_1 m_2}\right.
\times\big[\frac{\kappa \kappa_{1}}{|m_2|}\sqrt{\frac{(l_{2}+1-|m_2|)!(l_{1}+|m_2|)!}{(l_{2}+1+|m_2|)!(l_{1}-|m_2|)!}} \nonumber\\
+\left.\frac{\kappa \kappa_{-1}}{|m_2|}\sqrt{\frac{(l_{2}-1-|m_2|)!(l_1+|m_2|)!}{(l_{2}-1+|m_2|)!(l_1-|m_2|)!}}\big]\right) & \mbox{if ($l_1 <l_2$ and $l_2\geq2$)} \nonumber\\ \\
\end{array}
\right.
\end{eqnarray}

To avoid any confusion, we reiterate that the above results hold for
PT-scan approximation and a NC-beam function with discrete even-fold
azimuthal symmetry. Other residual symmetries in NC-beam can reduce
the set of non-zero beam BipoSH further.  In particular, if the
experimental beam has reflection symmetry, then odd parity beam BipoSH
will vanish and only even parity ones will be present. This implies
that odd parity beam BipoSH can be used as a measure of breakdown of
reflection symmetry in NC-beams.

\bibliographystyle{JHEP}
\bibliography{reference}

\end{document}